\documentclass[prd,aps,tightenlines,a4paper,12pt]{revtex4}

\usepackage{graphicx}
\usepackage{bm}
\usepackage{url}
\usepackage{amsfonts}
\usepackage{amsmath}

\newcommand{\ve}[1]{\mbox{\boldmath$#1$}}

\begin{document}

\title{A detailed proof of the fundamental theorem of STF multipole expansion in linearized gravity}

\author{Sven \surname{Zschocke}}
\affiliation{
Lohrmann Observatory, Dresden Technical University,\\
Helmholtzstrasse 10, D-01069 Dresden, Germany\\
}

\begin{abstract}


The linearized field equations of general relativity in harmonic coordinates are given by an inhomogeneous wave equation. 
In the region exterior to the matter field, the retarded solution of this wave equation can be expanded in terms of 
$10$ Cartesian symmetric and tracefree (STF) multipoles in post-Minkowskian approximation. 
For such a multipole decomposition only three and rather weak assumptions are required:
\begin{enumerate}
\item[] 
\begin{enumerate}
\item[1.] No-incoming radiation condition.
\item[2.] The matter source is spatially compact.
\item[3.] A spherical expansion for the metric outside the matter source is possible.
\end{enumerate}
\end{enumerate}

\noindent
During the last decades, the STF multipole expansion has been established as a powerful tool 
in several fields of gravitational physics: celestial mechanics, theory of gravitational waves and 
in the theory of light propagation and astrometry. But despite its formidable importance, an explicit proof of 
the fundamental theorem of STF multipole expansion has not been presented thus far, while only some parts of 
it are distributed into several publications. In a technical but more didactical form, an explicit 
and detailed mathematical proof of each individual step of this important theorem of STF multipole expansion 
is represented. 

\end{abstract}


\maketitle

\newpage

\tableofcontents

\newpage

\section{Introduction}

The field equations of gravity, {\it Einstein} (1915,1916) \cite{Einstein1,Einstein2}, constitute a set of $10$ coupled 
nonlinear partial differential equations which relate the metric tensor $g^{\alpha \beta}$ of curved space-time 
 to the stress-energy tensor of matter $T^{\alpha\beta}$. Due to the inherited mathematical 
difficulties of solving these field equations in closed form, exact and physically well interpretable solutions of 
general theory of relativity are on rare occasions \cite{Book_Exact_Solutions}. The most well-known examples for the 
case of massive isolated sources are the metric of a spherically symmetric massive body derived 
by {\it Schwarzschild} (1916) \cite{Schwarzschild}, the solution for a spherically symmetric and electrically charged 
body found by {\it Reissner} (1916) \cite{Reissner} and {\it Nordstr\"om} (1918) \cite{Nordstrom}, and the metric for 
rotating bodies obtained by {\it Kerr} (1963) \cite{Kerr}. However, for more realistic scenarios, like an accelerated body, 
an asymmetric body, or a N-body system, exact solutions for the metric field are far out of reach or 
even do not exist. Therefore, approximative 
approaches of general relativity are essential for further progress in the theory of gravity. An important approximative 
approach is the theory of linearized gravity, where in harmonic gauge the coupled field equations of {\it Einstein's} 
theory are simplified to a set of decoupled inhomogeneous wave equations for each of the $10$ components of the 
metric tensor, {\it Einstein} (1916) \cite{Einstein3}:
\begin{eqnarray}
\square_x \overline{h}^{\alpha\beta} \left(t , \ve{x}\right) &=&
- \frac{16\,\pi\,G}{c^4}\;T^{\alpha\beta} \left(t , \ve{x}\right),
\label{Introduction_1}
\end{eqnarray}

\noindent
which is valid up to order ${\cal O} \left(G^2\right)$, and  $G$ is the gravitational constant. 
In Eq.~(\ref{Introduction_1}), $\displaystyle \square_x$ is the d'Alembert operator, 
$\overline{h}^{\alpha\beta} = \eta^{\alpha \beta} - \sqrt{-g}\,g^{\alpha \beta}$ is the metric perturbation 
($g=$ determinant of $g^{\alpha \beta}$, $\eta^{\alpha \beta} = {\rm diag} \left(-1,+1,+1,+1\right)$ is the 
metric of flat space-time), and $c$ is the speed of light; the curved spacetime is assumed to be covered 
by harmonic coordinates $\left(t,\ve{x}\right)$. 

The mathematical structure of linearized field equations (\ref{Introduction_1}) resembles the field equations of 
classical electrodynamics in Lorentz gauge, $\square_x A^{\mu} = - \frac{\displaystyle 4\,\pi}{\displaystyle c}\,j^{\mu}$, 
with $A^{\mu}$ being the four-potential and $j^{\mu}$ the four-current, but with the addition that in classical 
electrodynamics the space-time is Minkowskian, while the space-time in linearized gravity is in fact 
curved.   
Especially, the Green functions of both field equations are formally the same, and the harmonic 
coordinates $\left(t,\ve{x}\right)$ are treated as though they were Cartesian coordinates in flat 
Minkowski space \cite{Thorne}. 
Hence, like in classical electrodynamics, a solution of (\ref{Introduction_1}) is given by,  
{\it Einstein} (1916) \cite{Einstein3}: 
\begin{eqnarray}
\overline{h}^{\alpha \beta} \left(t,\ve{x}\right) &=& \frac{4\,G}{c^4}\,
\int_V d^3 x^{\prime}\, 
\frac{T^{\alpha\beta}\left(t_{\rm ret}, \ve{x}^{\prime}\right)}{\left| \ve{x}^{\prime} - \ve{x} \right|}\,,
\label{Introduction_2}
\end{eqnarray}

\noindent
where the integral runs over some finite spatial volume $V$ of the extended matter field, 
$t_{\rm ret} = t - \frac{\displaystyle \left| \ve{x}^{\prime} - \ve{x} \right|}{\displaystyle c}$ is the retarded 
time from a point inside the matter source with spatial coordinate $\ve{x}^{\prime}$ to a field point with spatial 
coordinate $\ve{x}$. The so-called advanced solution, where $t_{\rm ret}$ in (\ref{Introduction_2}) is replaced by 
$t_{\rm adv} = t + \frac{\displaystyle \left| \ve{x}^{\prime} - \ve{x} \right|}{\displaystyle c}$, is usually be regarded 
unphysical because it violates the causality condition and will not be considered here. 

The multipole decomposition of (\ref{Introduction_2}) in terms of spherical harmonics is a highly effective approach 
to further analyze this solution. That tool of multipole expansion has originally been applied  
a long time ago in classical electrodynamics \cite{Jackson_Electrodynamics} and later been transformed into the 
case of linearized gravity. In this respect, a bench mark was the investigation 
of {\it Campbell}, {\it Macek} $\&$ {\it Morgan} (1977) \cite{CMM} who have worked out a multipole decomposition of the 
scalar (gravitational potential), vectorial (electrodynamical four-potential) and tensorial (linearized gravity) field 
outside the matter source in terms of spherical harmonics.

However, the use of so-called Cartesian symmetric and tracefree (STF) multipole moments 
\cite{Sachs,Pirani,Courant_Hilbert,Gelfand,Coope1,Coope2,Coope3} instead of spherical harmonics simplifies considerably
the calculations in gravitational physics \cite{Thorne,Blanchet_Damour1,Blanchet_Damour2,Hartmann_Soffel_Kioustelidis}: 
the mathematical relations and expressions in gravitational theory become simpler, the numerical algorithms 
can be performed more efficiently, and the whole approach of gravitational theory becomes more elegant. 
By now, the STF multipole expansion, in post-Newtonian approximation ("weak-field slow-motion approximation", i.e.  
$g_{00}, g_{ij}$ exact to order ${\cal O} \left(c^{-2}\right)$, $g_{0i}$ exact to order ${\cal O} \left(c^{-3}\right)$) 
and post-Minkowskian approximation ("weak-field approximation", i.e. $g_{\alpha\beta}$ exact to order ${\cal O} \left(G\right)$), has been established 
as an important tool in linearized gravity and has found a wide range of applications: in celestial mechanics 
\cite{Hartmann_Soffel_Kioustelidis,DSX1,DSX2}, in the theory of gravitational waves 
\cite{Eubanks,Radiation_Condition,Gravitational_Waves2}, and in the theory of light propagation in curved 
space-time \cite{Lightpropagation1,Lightpropagation2,Lightpropagation3,Lightpropagation4} which is a fundamental 
aspect of relativistic astrometry. Meanwhile, the STF multipole expansion in linearized gravity has a remarkable history 
and encompasses some decades of period of time. Let us mention some important contributions which are considered as 
cornerstones in the theory of multipole expansion; further historical facts can be found, for instance, in Box $1$ 
in \cite{Thorne}, introductory sections in \cite{Blanchet_Damour1} and \cite{History_MP}, and in Section $4.4$ 
in \cite{Kopeikin_Book}.

First, the approach developed in \cite{CMM} has been established in terms of STF tensors in a pioneering work 
by {\it Thorne} (1980) \cite{Thorne} in post-Newtonian approximation, where  some first steps of earlier 
investigations \cite{Sachs,Pirani,Epstein_Wagoner,Wagoner} have considerably been generalized. Especially, 
{\it Thorne} (1980) \cite{Thorne} has shown that the metric outside the matter source can be expanded 
in terms of $10$ STF tensors as follows (Eqs.~(8.4) in \cite{Thorne}): 
\begin{eqnarray}
\overline{h}^{\alpha \beta} \left(t , \ve{x}\right) &=&
\frac{4\,G}{c^4}\;\sum\limits_{l=0}^{\infty} \frac{\left(-1\right)^l}{l!}\,\partial_L\,
\left[\frac{\hat{F}_L^{\alpha \beta} \left(u\right)}{r}\right]\,,
\label{Introduction_3}
\end{eqnarray}

\noindent
where $\partial_L = \frac{\displaystyle \partial^l}{\displaystyle \partial x^{a_1}\,...\,\partial x^{a_l}}$ are
$l$ spatial derivatives, $r = \left|\ve{x}\right|$ is the spatial distance between the origin of coordinate system and 
the field point with spatial coordinate $\ve{x}$, $\hat{F}_L^{\alpha\beta}$ are $10$ STF multipoles, and $u = c t - r$. 
Moreover, {\it Thorne} (1980) \cite{Thorne} has shown, using energy-momentum conservation (Eqs.~(8.6) and (8.7) 
in \cite{Thorne}) and a sophisticated gauge transformation (Eqs.~(8.9) in \cite{Thorne}) which preserves the harmonic 
gauge, that outside the matter the metric can finally be expressed in terms of $2$ independent multipoles in 
post-Newtonian approximation: mass multipoles $\hat{M}_L$ and spin multipoles $\hat{S}_L$ (Eqs.~(8.13) in \cite{Thorne}). 
However, the multipoles (Eqs.~(5.32) in \cite{Thorne}) were still formally divergent at spatial infinity. 

Consequently, {\it Blanchet} $\&$ {\it Damour} (1986) \cite{Blanchet_Damour1} have further developed the approach 
in \cite{Thorne} and have demonstrated that {\it Thorne}'s post-Newtonian multipoles are physically meaningful if one 
makes a rigorous use of the compact-support source of energy-momentum tensor. 
This important result has been achieved with the aid of the theory 
of distributions by means of which {\it Blanchet} $\&$ {\it Damour} (1986) \cite{Blanchet_Damour1} 
were able to extract the physically relevant and non-divergent part of {\it Thorne}'s multipoles in post-Newtonian 
approximation. 

Finally, {\it Blanchet} $\&$ {\it Damour} (1989) \cite{Blanchet_Damour2} have established a powerful theorem in 
post-Minkowskian approximation which states that outside of an isolated source the metric can be expanded in terms of 
$10$ Cartesian STF multipoles (\ref{Introduction_3}) (Eqs.~(B.2) and (B.3) in \cite{Blanchet_Damour2}),  
defined by 
\begin{eqnarray}
\hat{F}_L^{\alpha \beta} \left(u\right) &=& \int_V d^3 x^{\prime}\;\hat{x}_L^{\prime}\int\limits_{-1}^{+1} d z\;
\delta_l \left(z\right)\;T^{\alpha \beta} \left(\frac{u + z\,r^{\prime}}{c}, \ve{x}^{\prime}\right),
\label{Introduction_4}
\end{eqnarray}

\noindent
where $\hat{x}^{\prime}_L = \underset{a_1\,,...\,,a_l}{\rm STF}\left(x^{\prime}_{a_1}\,,...\,,x^{\prime}_{a_l}\right)$, 
and $r^{\prime} = \left|\ve{x}^{\prime}\right|$ is the distance between the origin of coordinate system and a point 
inside of the source with spatial coordinate $\ve{x}^{\prime}$; the coefficient functions in (\ref{Introduction_4}) are 
given by
\begin{eqnarray}
\delta_l (z) &=& \frac{\left(2\,l + 1\right)!!}{2^{l+1}\,l!}\;\left(1 - z^2\right)^l\,,
\label{Introduction_5}
\end{eqnarray}

\noindent
which are normalized: $\int\limits_{-1}^{+1} d z\,\delta_l \left(z\right)=1$. 
As we will see, the expansion (\ref{Introduction_3}) is valid in regions $r > r_0$, where $r_0$ is the 
radius of the smallest possible sphere which contains completely the source of matter.    
The 
expansion (\ref{Introduction_3}) - (\ref{Introduction_4}) represents the fundamental theorem of STF multipole 
expansion in linearized gravity, e.g. Eqs.~(B.2) - (B.3) in \cite{Blanchet_Damour2},  
Eqs.~(5.3) - (5.4) in \cite{Multipole_Damour_2}, Eqs.~(56) - (57) in \cite{Radiation_Condition}, 
or Eq.~(25) in \cite{Gravitational_Waves2}, and stands for a 
solution of linearized field equations (\ref{Introduction_1}) in post-Minkowskian approximation, hence it is even valid 
in case of ultra-relativistic motion of matter inside the source. 

After all, using energy-momentum conservation (Eqs.~(5.14) and (5.18) in \cite{Multipole_Damour_2}) and applying 
a sophisticated gauge choice (Eq.~(5.31) in \cite{Multipole_Damour_2}) {\it Damour} $\&$ {\it Iyer} (1991) 
\cite{Multipole_Damour_2} have demonstrated, footing on the pioneering works of {\it Thorne} (1980) \cite{Thorne} and 
{\it Blanchet} $\&$ {\it Damour} (1986,1989) \cite{Blanchet_Damour1,Blanchet_Damour2}, that also in post-Minkowskian 
approximation the family of these $10$ multipoles can be reduced to finally only $2$ independent multipoles: mass 
multipoles $\hat{M}_L$ (Eq.~(5.33) in \cite{Multipole_Damour_2}) and spin multipoles $\hat{S}_L$ 
(Eq.~(5.35) in \cite{Multipole_Damour_2}):
\begin{eqnarray}
\overline{h}^{\alpha \beta} &=& \overline{h}^{\alpha \beta} \left(\hat{M}_L\,,\,\hat{S}_L\right).
\label{Introduction_6}
\end{eqnarray}

\noindent
The demonstration, that the metric in (\ref{Introduction_3}) which depends on $10$ multipoles $\hat{F}_L^{\alpha\beta}$ can 
be reduced to the form in (\ref{Introduction_6}) where the metric depends only on $2$ multipoles 
$(\hat{M}_L,\hat{S}_L)$, is a rather ambitious assignment of a task and makes extensive use of irreducible Cartesian tensor 
techniques originally introduced in \cite{Coope1,Coope2,Coope3}. 
{\it Damour} $\&$ {\it Iyer} (1991) \cite{Multipole_Damour_2} have also demonstrated that to order 
${\cal O} \left(c^{-4}\right)$ their post-Minkowskian multipoles coincide with the post-Newtonian multipoles 
of {\it Blanchet} $\&$ {\it Damour} (1989) \cite{Blanchet_Damour2} (Eqs.~(5.38) and (5.41) in \cite{Multipole_Damour_2}).
So the investigation in \cite{Multipole_Damour_2} has been the final touch in the approach of STF multipole expansion to 
order ${\cal O} \left(G\right)$. This elaborated work of {\it Damour} $\&$ {\it Iyer} (1991) \cite{Multipole_Damour_2} will, 
however, not be on the scope of the present investigation. Instead, we will be focussed on 
theorem (\ref{Introduction_3}) - (\ref{Introduction_4}), which is the heart and the core part of STF multipole 
expansion. An explicit proof of this important theorem is not so 
straightforward as one might believe and has not been presented in detail thus far; only some parts of it are published 
but scattered in several publications \cite{CMM,Thorne,Blanchet_Damour1,Blanchet_Damour2}. 
Here, in view of its formidable relevance in the theory of linearized gravity, we will outline a more detailed 
mathematical proof of each individual step of multipole expansion (\ref{Introduction_3}) in post-Minkowskian approximation. 

The paper is organized as follows: In section \ref{section_A} a compendium of the exact field equations of gravity is 
provided. The linearized approximation of general relativity is given in section \ref{section_B}. 
Section \ref{section_C} is devoted to the main part of our investigation, where a detailed proof of the fundamental 
theorem (\ref{Introduction_3}) is represented and the required assumptions for its validity are defined. A summary is 
finally given in section \ref{section_D}. 

We shall use fairly standard notations of the STF tensor approach 
\cite{Pirani,Thorne,Blanchet_Damour1,Multipole_Damour_2,Hartmann_Soffel_Kioustelidis}:

\begin{itemize}

\item Lower case Latin indices $i$, $j$, \dots take values $1,2,3$.
\item Lower case Greek indices $\mu$, $\nu$, \dots take values $0,1,2,3$.
\item $\delta_{ij} = \delta^{ij} = {\rm diag} \left(+1,+1,+1\right)$ is Kronecker delta.
\item $n! = n \left(n-1\right)\left(n-2\right)\cdot\cdot\cdot 2 \cdot 1$ is the faculty for positive integer;  
$0! = 1$. 
\item $n!! = n \left(n-2\right) \left(n-4\right)\cdot\cdot\cdot \left(2\;{\rm or}\;1\right)$ is the double faculty 
for positive integer; $0!! = 1$. 
\item $L=i_1 i_2 ...i_l$ and $Q=i_1 i_2 ...i_q$ are Cartesian multi-indices of a given tensor $T$, that means  
$T_L \equiv T_{i_1 i_2 \,.\,.\,.\,i_l}$ and  $T_Q \equiv T_{i_1 i_2 \,.\,.\,.\,i_q}$, respectively. 
\item two identical multi-indices imply summation: 
$A_L\,B_L \equiv \sum\limits_{i_1\,.\,.\,.\,i_l}\,A_{i_1\,.\,.\,.\,i_l}\,B_{i_1\,.\,.\,.\,i_l}$.
\item The symmetric part of a Cartesian tensor $T_L$ is, cf. Eq.~(2.1) in \cite{Thorne}: 
\begin{eqnarray} 
T_{\left(L\right)} &=& T_{\left(i_1 ... i_l \right)} = \frac{1}{l!} \sum\limits_{\sigma} 
A_{i_{\sigma\left(1\right)} ... i_{\sigma\left(l\right)}}\,,
\end{eqnarray}

\noindent
where $\sigma$ is running over all permutations of $\left(1,2,...,l\right)$.

\item The symmetric tracefree part of a Cartesian tensor $T_L$ (notation: 
$\hat{T}_L \equiv \underset{L}{\rm STF}\,T_L$) is, cf. Eq.~(2.2) in \cite{Thorne}:
\begin{eqnarray}
\hat{T}_L &=& \sum_{k=0}^{\left[l/2\right]} a_{l k}\,\delta_{(i_1 i_2 ...} \delta_{i_{2k-1} i_{2k}}\,
S_{i_{2k+1 ... i_l) \,a_1 a_1 ... a_k a_k}}\,,
\label{anti_symmetric_1}
\end{eqnarray}

\noindent
where $\left[l/2\right]$ means the largest integer less than or equal to $l/2$, and $S_L \equiv T_{\left(L\right)}$ 
abbreviates the symmetric part of tensor $T_L$. For instance, $T_L^{\alpha\beta}$ means STF with respect to indices $L$ 
but not with respect to indices $\alpha,\beta$. The coefficient in (\ref{anti_symmetric_1}) is given by 
\begin{eqnarray}
a_{l k} &=& \left(-1\right)^k \frac{l!}{\left(l - 2 k\right)!}\,
\frac{\left(2 l - 2 k - 1\right)!!}{\left(2 l - 1\right)!! \left(2k\right)!!}\,.
\label{coefficient_anti_symmetric}
\end{eqnarray}

\noindent
As instructive examples of (\ref{anti_symmetric_1}) let us consider the cases $l=2$ and $l=3$:
\begin{eqnarray}
\hat{T}_{ij} &=& T_{\left(ij\right)} - \frac{1}{3}\,\delta_{ij}\,T_{ss}\,,
\label{anti_symmetric_2}
\\
\nonumber\\
\hat{T}_{ijk} &=& T_{\left(ijk\right)} - \frac{1}{5} 
\left(\delta_{ij} T_{\left(kss\right)} + \delta_{jk} T_{\left(iss\right)} + \delta_{ki} T_{\left(jss\right)}\right).
\label{anti_symmetric_3}
\end{eqnarray}

\end{itemize}

Further STF relations can be found in \cite{Thorne,Blanchet_Damour1}.
We also will make use of {\it Einstein's} sum convention, that mans repeated indices are implicitly summed over. 

\section{Einstein's field equations}\label{section_A}

The gravitation is described by $10$ coupled nonlinear partial differential equations 
for the metric tensor, {\it Einstein} (1915,1916) \cite{Einstein1,Einstein2}, which can be written in the form:
\begin{eqnarray}
R^{\alpha\beta} - \frac{1}{2}\,g^{\alpha\beta}\,R &=& \frac{8\,\pi\,G}{c^4}\,T^{\alpha\beta}\,,
\label{Field_Equations_5}
\end{eqnarray}

\noindent
and which discover a fundamental relation between the metric of space-time and the matter 
field. Essentially, (\ref{Field_Equations_5}) represents a relation among contravariant tensors, of which $R^{\alpha\beta}$ is the Ricci 
curvature tensor, $g^{\alpha\beta}$ is the metric tensor with signature $\left(-\,,\,+\,,\,+\,,\,+\right)$, and 
$T^{\alpha\beta}$ is the energy-momentum tensor of matter; $R = R^{\alpha}_{\alpha}$ is the Ricci scalar of curvature. The 
field equations (\ref{Field_Equations_5}) are valid in any coordinate system, that means the 
coordinates are still arbitrary. For an asymptotically flat space-time, it is useful to decompose the  
metric tensor as follows:
\begin{eqnarray}
\sqrt{-g}\,g^{\alpha\beta} &=& \eta^{\alpha \beta} - \overline{h}^{\alpha \beta}\,,
\label{metric_20}
\end{eqnarray}

\noindent
where $g$ is the determinant of metric tensor $g^{\alpha\beta}$, $\overline{h}^{\alpha \beta}$ is the metric 
perturbation which describes the deviation of the metric tensor of curved space-time from the metric tensor of 
Minkowskian flat space-time given by  
\begin{eqnarray}
\eta_{\alpha\beta} &=& \eta^{\alpha\beta} = {\rm diag} \left(-1\,,\,+1\,,\,+1\,,\,+1\right). 
\label{metric_10}
\end{eqnarray}

\noindent
In harmonic gauge, also known as de Donder gauge, 
\begin{eqnarray}
\partial_{\beta}\;\overline{h}^{\alpha \beta} &=& 0\,,
\label{metric_30}
\end{eqnarray}

\noindent
{\it Einstein's} field equations (\ref{Field_Equations_5}) read (Eq.~(36.37) in \cite{MTW}, Eq.~(5.2b) in \cite{Thorne}):
\begin{eqnarray}
\square_x \overline{h}^{\alpha \beta} &=& - \frac{16\,\pi\,G}{c^4}\,\left(\tau^{\alpha \beta} + t^{\alpha \beta}\right),
\label{Field_Equations_10}
\end{eqnarray}

\noindent
where 
$\square_x = \eta^{\mu\nu}\,
\frac{\displaystyle \partial}{\displaystyle\partial x^{\mu}}\,\frac{\displaystyle\partial}{\displaystyle\partial x^{\nu}}$  
is the d'Alembert operator. These both tensors in (\ref{Field_Equations_10}) are given by 
(Eq.~(5.3) in \cite{Thorne})
\begin{eqnarray}
\tau^{\alpha \beta} &=& \left( - g\right)\,T^{\alpha \beta}\,,
\label{metric_35}
\\
\nonumber\\
t^{\alpha \beta} &=& \left( - g\right)\,t_{\rm LL}^{\alpha \beta} + \frac{c^4}{16\,\pi\,G}\;
\left(\overline{h}^{\alpha\mu}_{\;\;\;,\;\nu}\;\overline{h}^{\beta \nu}_{\;\;\;,\;\mu} 
- \overline{h}^{\alpha\beta}_{\;\;\;,\;\mu\nu}\;\overline{h}^{\mu \nu}\right),
\label{metric_40}
\end{eqnarray}

\noindent
where $t_{\rm LL}^{\alpha \beta}$ is the Landau-Lifschitz pseudotensor of gravitational field, in explicit form 
given by Eq.~(20.22) in \cite{MTW} or by Eqs.~(96.8) and (96.9) in \cite{Landau_Lifschitz}. The field 
equations (\ref{Field_Equations_10}) are exact and the gravitational field is not necessarily weak, because the only 
assumptions made so far are the decomposition (\ref{metric_20}) and the choice of a harmonic coordinate system 
(\ref{metric_30}).

\section{Linearized theory of gravity and STF multipole expansion}\label{section_B}

The tensors (\ref{metric_35}) and (\ref{metric_40}) can be expanded in terms of the 
coupling constant, cf. Eq.~(3.528) and Eq.~(3.529) in \cite{Kopeikin_Book}:
\begin{eqnarray}
\tau^{\alpha\beta} &=& T^{\alpha\beta} + \frac{8\,\pi\,G}{c^4}\,\tau_1^{\alpha\beta} + {\cal O} \left(G^2\right),
\label{solution_20}
\\
\nonumber\\
t^{\alpha\beta} &=& \frac{8\,\pi\,G}{c^4}\,t_1^{\alpha\beta} + {\cal O} \left(G^2\right).
\label{solution_25}
\end{eqnarray}

\noindent
As it can be deduced from Eq.~(\ref{Field_Equations_10}) and expansions (\ref{solution_20}) - (\ref{solution_25}), 
in harmonic gauge and up to terms of the order ${\cal O} \left(G^2\right)$, the Einstein's field equations 
are simplified to d'Alembert's wave equation for each of the $10$ components of the metric tensor, 
{\it Einstein} (1916) \cite{Einstein3}:   
\begin{eqnarray}
\square_x \overline{h}^{\alpha\beta} \left(t , \ve{x}\right) &=&
- \frac{16\,\pi\,G}{c^4}\;T^{\alpha\beta} \left(t , \ve{x}\right),
\label{metric_A}
\end{eqnarray}

\noindent
which is called linearized gravity, a term which refers to the fact that the approximative field equations (\ref{metric_A}) 
are linear partial differential equations, to be contrary to the nonlinear exact field equations of 
gravity (\ref{Field_Equations_5}) or (\ref{Field_Equations_10});    
recall the harmonic coordinates are denoted by $\left(t,\ve{x}\right)$. 

Actually, there are formally infinitely many solutions of wave equation (\ref{metric_A}). These solutions 
of (\ref{metric_A}) consist of a general solution of the homogeneous wave 
equation $\square_x \overline{h}_{\rm hom}^{\alpha\beta} = 0$ plus one particular solution of inhomogeneous wave equation 
(\ref{metric_A}): 
$\overline{h}^{\alpha\beta} = \overline{h}^{\alpha\beta}_{\rm hom} + \overline{h}^{\alpha\beta}_{\rm inhom}$. 
For an unique solution of (\ref{metric_A}) one has to impose initial and boundary conditions. In case of an infinite 
space-time, there are no boundary conditions, and a well-posed problem (i.e. existence of one and only one unique solution) 
is given by the initial value problem at initial time $t^{\prime\prime}$ (Cauchy problem):
\begin{eqnarray}
\overline{h}_{\rm hom}^{\alpha\beta} \left(t^{\prime\prime},\ve{x}^{\prime\prime}\right), \quad
\frac{\partial}{\partial c\,t^{\prime\prime}}\,\overline{h}_{\rm hom}^{\alpha\beta}
\left(t^{\prime\prime},\ve{x}^{\prime\prime}\right).
\label{Cauchy_1}
\end{eqnarray}

\noindent
These initial conditions are valid in the entire three-dimensional space. According to {\it Kirchhoff's} rigorous 
integration of the wave equation, {\it Kirchhoff} (1883) \cite{Kirchhoff1}, an unique solution 
of (\ref{metric_A}) - (\ref{Cauchy_1}) is given in terms of these initial conditions by an integral over an arbitrarily 
shaped but sufficiently smooth surface which 
contains completely the field point $\ve{x}$ and the spatially compact matter field described by the energy-momentum tensor 
$T^{\alpha\beta}$; an explicit expression of {\it Kirchhoff's} solution can be found, for instance, 
in Eq.~(13) on page $420$ in \cite{Born}. Here, without loss of generality, the surrounding surface $\partial S$ is assumed 
to be the surface of a sphere $S$. Then, the unique solution of (\ref{metric_A}) - (\ref{Cauchy_1}) can be written as 
follows: 
\begin{eqnarray}
\overline{h}^{\alpha\beta} \left(t , \ve{x}\right) &=& \overline{h}^{\alpha\beta}_{\rm hom} \left(t,\ve{x}\right) 
+ \overline{h}^{\alpha\beta}_{\rm inhom} \left(t,\ve{x}\right), 
\label{solution_2a}
\\
\nonumber\\
\overline{h}^{\alpha\beta}_{\rm hom} \left(t,\ve{x}\right) &=& \frac{1}{4\,\pi} \int_{\partial S} d \Omega^{\prime\prime}
\left[\frac{\partial}{\partial r^{\prime\prime}} \left(r^{\prime\prime}\,\overline{h}_{\rm hom}^{\alpha\beta}
\left(t^{\prime\prime},\ve{x}^{\prime\prime}\right) \right) +
\frac{\partial}{\partial c t^{\prime\prime}} \left(r^{\prime\prime}\,\overline{h}_{\rm hom}^{\alpha\beta}
\left(t^{\prime\prime},\ve{x}^{\prime\prime}\right)\right)\right],
\label{solution_2b}
\\
\nonumber\\
\overline{h}^{\alpha\beta}_{\rm inhom} \left(t,\ve{x}\right) &=& \frac{4\,G}{c^4}\;\int\limits_{-\infty}^{t} d t^{\prime}
\int_V d^3 x^{\prime}\; T^{\alpha\beta}\left(t^{\prime}, \ve{x}^{\prime}\right)
\frac{\delta \left( t^{\prime} - t + \frac{\displaystyle \left| \ve{x}^{\prime} - \ve{x} \right|}{\displaystyle c}\right)}
{\left| \ve{x}^{\prime} - \ve{x} \right|}\,.
\label{solution_2c}
\end{eqnarray}

\begin{figure}[!h]
\begin{center}
\includegraphics[scale=0.3]{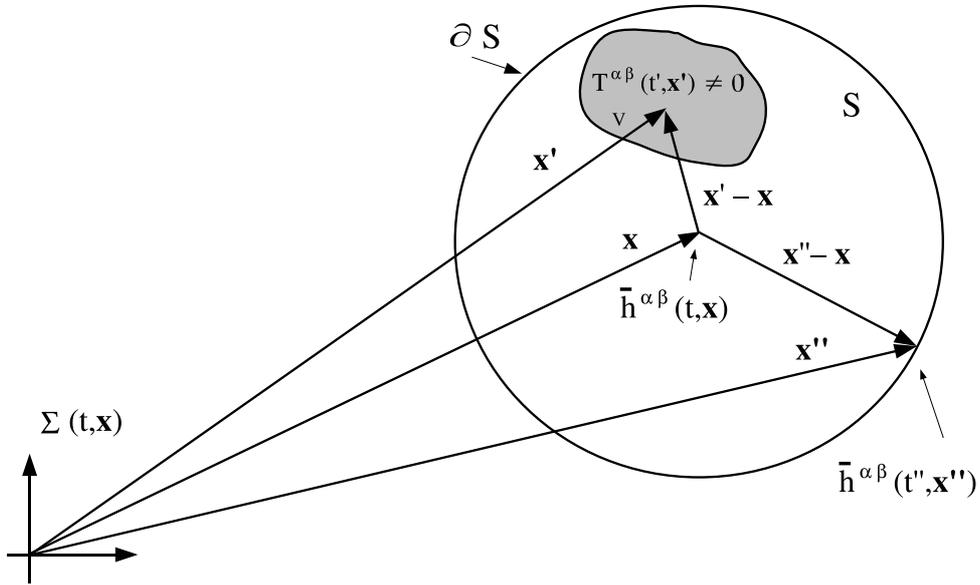}
\caption{Graphical representation of {\it Kirchhoff's} solution (\ref{solution_2a}) - (\ref{solution_2c}). 
The global reference frame with 
harmonic coordinates $\left(t,\ve{x}\right)$ is denoted by $\Sigma \left(t,\ve{x}\right)$. The vector $\ve{x}$ points from 
the origin of global coordinate system to the field point with spatial coordinate $\ve{x}$. The matter field is described 
by an energy-momentum tensor $T^{\alpha \beta}$. The matter field and the field point are enclosed by a virtual sphere $S$ 
with surface $\partial S$. The center of the sphere is located at spatial coordinate $\ve{x}$ of the field point, so that 
the spatial distance between the field point and a point inside the matter field is given by 
$\left|\ve{x}^{\prime}-\ve{x}\right|$, and the spatial distance between the field point and the surface $\partial S$ is 
given by $\left|\ve{x}^{\prime\prime}-\ve{x}\right|$. The matter field is assumed to be isolated, that means: 
$1.$ outside the 
region of some finite spatial volume $V$ (gray colored) the matter field vanishes, and $2.$ there is no gravitational 
radiation from outside through the surface $\partial S$ of sphere $S$. The metric field $\overline{h}^{\alpha \beta}$ at 
field point $\left(t,\ve{x}\right)$ and at surface point $\left(t^{\prime\prime},\ve{x}^{\prime\prime}\right)$ has also 
been indicated.}
\label{FIG:Fig_4}
\end{center}
\end{figure}

\noindent
Here, $\delta \left(x\right)$ is Dirac's delta-distribution,
normalized by $\int\limits_{- \infty}^{+ \infty} d x\,\delta\left(x\right) = 1$.
For a graphical elucidation of Eqs.~(\ref{solution_2a}) - (\ref{solution_2c}) see Fig.~\ref{FIG:Fig_4}. 
According to (\ref{solution_2b}), the solution of homogeneous wave equation is given by a surface integral over a sphere, 
while (\ref{solution_2c}) is the particular solution of inhomogeneous wave equation which is called retarded solution. 
In (\ref{solution_2b}) we use $r^{\prime\prime} = \left|\ve{x}^{\prime\prime} - \ve{x}\right|$, and for the retarded time 
between field point $\ve{x}$ and any point $\ve{x}^{\prime\prime}$ on the surface of sphere we use 
$t^{\prime\prime} = t - \frac{\displaystyle r^{\prime\prime}}{\displaystyle c}$. 
The homogeneous solution (\ref{solution_2b}) contains the initial conditions (\ref{Cauchy_1}), that means 
$\overline{h}^{\alpha\beta}_{\rm hom} \left(t,\ve{x}\right)$ in the whole space-time is uniquely determined by its 
initial values (\ref{Cauchy_1}) on surface $\partial S$; cf. Eq.~(9) in \cite{Radiation_Condition}. The surface integral 
is given in terms of spherical coordinates $\left(r^{\prime\prime},\theta^{\prime\prime},\phi^{\prime\prime}\right)$ and 
the origin of the spherical coordinate system is located at the center of the sphere $S$, so that 
$d \Omega^{\prime\prime} = \sin \theta^{\prime\prime}\,d \theta^{\prime\prime}\,d \phi^{\prime\prime}$. 
The integration in (\ref{solution_2b}) runs over the surface with radius $r^{\prime\prime}$. 

\noindent
Physically, {\it Kirchhoff's} theorem \cite{Kirchhoff1} states that the homogeneous solution (\ref{solution_2b}) is 
uniquely determined by source points which form a sphere with arbitrarily large radius $r^{\prime\prime}$. 
We will assume that the matter source $T^{\alpha \beta}$ in (\ref{metric_A}) is isolated, that means the source is spatially 
compact and does not receive any radiation from other sources far away; note, however, that the matter source itself 
can emit gravitational radiation. 
Accordingly, since the radius of sphere $S$ can be arbitrarily large, we are allowed to take the limit up to spatial 
infinity and can replace the initial conditions (\ref{Cauchy_1}) by the so-called no-incoming-radiation condition, 
cf. Eq.~(10) in \cite{Radiation_Condition}, and cf.~Eqs.~(2.5) and (2.6) in \cite{KlionerKopeikin1992}: 
\begin{eqnarray}
\lim_{\displaystyle r^{\prime\prime} \rightarrow + \infty \atop {\displaystyle t^{\prime\prime}} 
{\displaystyle \,+\,} \frac{\displaystyle r^{\prime\prime}}{\displaystyle c} {\displaystyle \,=\,} 
{\displaystyle \rm const.}} \left[\frac{\partial}{\partial r^{\prime\prime}}
\left(r^{\prime\prime}\,\overline{h}_{\rm hom}^{\alpha\beta}\left(\ve{x}^{\prime\prime},t^{\prime\prime}\right) \right) 
+ \frac{\partial}{\partial c t^{\prime\prime}} \left(r^{\prime\prime}\, \overline{h}_{\rm hom}^{\alpha\beta}
\left(\ve{x}^{\prime\prime},t^{\prime\prime}\right)\right)\right] &=& 0\,.
\label{No_Radiation_Condition}
\end{eqnarray}

\noindent
If we impose the no-incoming-radiation condition (\ref{No_Radiation_Condition}), then the unique solution 
of (\ref{metric_A}) is given by the retarded solution (\ref{solution_2c}), {\it Einstein} (1916) \cite{Einstein3}: 
\begin{eqnarray}
\overline{h}^{\alpha \beta} \left(t , \ve{x}\right) &=& \frac{4\,G}{c^4}\;\int\limits_{-\infty}^{t} d t^{\prime} 
\int_V d^3 x^{\prime}\; T^{\alpha\beta}\left(t^{\prime}, \ve{x}^{\prime}\right) 
\frac{\delta \left( t^{\prime} - t + \frac{\displaystyle \left| \ve{x}^{\prime} - \ve{x} \right|}{\displaystyle c}\right)}
{\left| \ve{x}^{\prime} - \ve{x} \right|}\,.
\label{metric_B}
\end{eqnarray}

\noindent
According to the fundamental theorem of STF multipole expansion, outside the matter field of an isolated source 
the retarded solution in (\ref{metric_B}) can be decomposed in terms of $10$ STF multipoles: 
Eqs.~(\ref{Introduction_3}) - (\ref{Introduction_4}). In what follows, we will present a detailed 
proof of key formulae of this STF multipole expansion.

\section{Proof of STF multipole expansion}\label{section_C}

The inhomogeneous wave equation (\ref{metric_A}) is valid for any component of the tensors $\overline{h}^{\alpha \beta}$ 
and $T^{\alpha\beta}$, so we consider the inhomogeneous wave equation just for one of the field components: 
\begin{eqnarray}
\square_x \overline{h} \left(t,\ve{x}\right) &=& - 4\,\pi\;T\left(t,\ve{x}\right)\,,
\label{Proof_Theorem_5}
\end{eqnarray}

\noindent
so that $\overline{h}$ stands either for $\overline{h}^{0 0}$, $\overline{h}^{0 i}$, or $\overline{h}^{i j}$, while 
$T$ stands either for $\frac{\displaystyle 4\,G}{\displaystyle c^4}\,T^{00}$, 
$\frac{\displaystyle 4\,G}{\displaystyle c^4}\,T^{0i}$, or $\frac{\displaystyle 4\,G}{\displaystyle c^4}\,T^{ij}$, 
respectively. As it has been discussed above, if the source is isolated (i.e. source is spatially compact and 
no-incoming radiation) then there exists one 
and only one solution of (\ref{Proof_Theorem_5}), namely (cf. Eq.~(\ref{metric_B})):
\begin{eqnarray}
\overline{h} \left(t,\ve{x}\right) &=& \int\limits_{- \infty}^{t} dt^{\prime} \int_V d^3 x^{\prime} \; 
T\left(t^{\prime},\ve{x}^{\prime}\right)\;
G_R \left(t^{\prime} - t, \ve{x}^{\prime} - \ve{x} \right),
\label{Proof_Theorem_10}
\end{eqnarray}

\noindent
where the spatial integration runs over the volume $V$ of the source, and the retarded Green function is given by 
\begin{eqnarray}
G_R \left(t^{\prime} - t, \ve{x}^{\prime} - \ve{x} \right) &=&
\frac{\delta \left( t^{\prime} - t + \frac{\displaystyle \left| \ve{x}^{\prime} - \ve{x} \right|}{\displaystyle c}\right)}
{\left| \ve{x}^{\prime} - \ve{x} \right|}\;.
\label{Green_Function}
\end{eqnarray}

\noindent
The assumption that the source in (\ref{Proof_Theorem_10}) is spatially compact is formulated 
as follows (cf. text above Eq.~(B1 a) in \cite{Blanchet_Damour2},  
cf. text above Eq.~(3.1) in \cite{Multipole_Damour_2}, or  
cf. text in section II A in \cite{Blanchet1})  
\begin{eqnarray}
T\left(t^{\prime},\ve{x}^{\prime}\right) &=& 0 \quad {\rm for}\; \left|\ve{x}^{\prime}\right| > r_0\;,
\label{compact_0}
\end{eqnarray}

\noindent
where $r_0$ is the radius of some sphere which contains completely the source.

Consider the retarded Green function in (\ref{Green_Function}), which can be expanded 
in a series of Legendre polynomials $P_q$, cf. Eqs.~(D1) and (D2a) in \cite{Blanchet_Damour1}: 
\begin{eqnarray}
G_R \left(t^{\prime} - t, \ve{x}^{\prime} - \ve{x}\right) &=&
\frac{\Theta \left(t - t^{\prime}\right) \;\Theta \left( 1 - \left| \nu \right|\right)}{2\,r\,r^{\prime}}
\sum\limits_{q=0}^{\infty} \frac{ \left(2\,q + 1\right)!!}{q!}\;
\hat{n}_Q \left(\phi^{\prime}\,,\,\theta^{\prime}\right)\;
\hat{n}_Q \left(\phi\,,\,\theta\right)\;P_q \left(\nu\right)\,,
\nonumber\\
\label{Proof_Theorem_15}
\end{eqnarray}

\noindent
where $r = \left|\ve{x}\right|$, $r^{\prime} = \left|\ve{x}^{\prime}\right|$, and 
\begin{eqnarray}
\Theta \left(s\right) &=& 0 \quad {\rm for}\quad s < 0 \;,
\nonumber\\
\Theta \left(s\right) &=& 1 \quad {\rm for}\quad s \ge 0 \;,
\label{Step_Function}
\end{eqnarray}

\noindent
is the Heaviside step function, and
\begin{eqnarray}
\nu &=& \frac{r^2 + {r^{\prime}}^2 - c^2 \left(t - t^{\prime}\right)^2}{2\,r\,r^{\prime}}\,,
\label{Proof_Theorem_20}
\end{eqnarray}

\noindent 
is the argument of the Legendre polynomial.

\vspace{1.0cm}

{\footnotesize
{\bf Proof 1:}
We will show the validity of Eq.~(\ref{Proof_Theorem_15}). Some parts of this proof have been presented in \cite{CMM} 
in terms of spherical harmonics, while here we present a proof in terms of STF tensors. 
The Legendre polynomials can be defined by (Rodrigues' formula, Eq.~(12.65) in \cite{Arfken_Weber})
\begin{eqnarray}
P_l \left(z\right) &=& \frac{1}{2^l\,l!}\,\frac{d^l}{d z^l} \left(z^2 - 1\right)^l\;,
\label{Expansion_1}
\end{eqnarray}

\noindent
and the normalization is (Eq.~(12.48) in \cite{Arfken_Weber})
\begin{eqnarray}
\int\limits_{-1}^{+1} dz\, P_n \left(z\right) P_m \left(z\right) &=&  \frac{2}{2 n + 1} \,\delta_{n m}\;.
\label{Expansion_2}
\end{eqnarray}

\noindent

Consider two directions given by two normalized vectors 
$\displaystyle \ve{n} = \frac{\ve{x}}{r}$,  $\displaystyle \ve{n}^{\prime} = \frac{\ve{x}^{\prime}}{r^{\prime}}$, 
with $r = \left|\ve{x}\right|$ and $r^{\prime} = \left|\ve{x}^{\prime}\right|$, 
and $\gamma$ is the angle between $\ve{n} \left(\theta,\phi\right)$ and 
$\ve{n}^{\prime} \left(\theta^{\prime}, \phi^{\prime}\right)$, that means $\cos\gamma = \ve{n} \cdot \ve{n}^{\prime}$; 
this angle satisfies the trigonometric identity: $\cos \gamma = \cos \theta\,\cos \theta^{\prime} +
\sin \theta\,\sin \theta^{\prime}\,\cos \left(\phi - \phi^{\prime}\right)$ (Eq.~(12.168) in \cite{Arfken_Weber}).
Then, let us consider a function $F \left(z, r, r^{\prime}\right)$ which depends on $z = \cos\gamma$. Further, we assume 
the function $F$ to be an element of Hilbert space $V=L^2$ given by 
$V:= L^2 \left( z = \left[ - 1 , + 1\right]\,;\, {\cal R}\right)$, that means the function $F$ is square-integrable over the 
surface of the unit sphere. Then, such a function $F$ can be expanded in terms of Legendre polynomials 
(Eq.~(12.49) in \cite{Arfken_Weber}):
\begin{eqnarray}
F \left(z, r, r^{\prime}\right) &=& \sum\limits_{l=0}^{\infty} P_l \left(z\right)\,
F_l \left(r, r^{\prime}\right)\,,
\label{Expansion_5}
\end{eqnarray}

\noindent
where the coefficients are given by (cf. Eq.~(12.50) in \cite{Arfken_Weber}; $x = \cos \gamma$),
\begin{eqnarray}
F_l \left(r, r^{\prime}\right) &=& \frac{2 l + 1}{2} \int\limits_{-1}^{+1} d x\;P_l \left(x\right) \,
F \left(x, r, r^{\prime}\right)\,.
\label{Expansion_10}
\end{eqnarray}

\noindent
The Legendre polynomial addition theorem states (Eq.~(8.189) or Eq.~(12.170) in \cite{Arfken_Weber},
for a detailed proof see chapter $12.8$ in \cite{Arfken_Weber}):
\begin{eqnarray}
P_l \left(z\right) &=& \frac{4\,\pi}{2 l + 1}\,\sum\limits_{m=-l}^{l}\,Y_{l m} \left(\theta,\phi\right)
\; Y_{l m}^{\ast} \left(\theta^{\prime}, \phi^{\prime}\right)\,,
\label{Expansion_15}
\end{eqnarray}

\noindent
where $Y_{l m}$ are the spherical harmonics as defined in \cite{Arfken_Weber}. 
By inserting (\ref{Expansion_10}) and (\ref{Expansion_15}) into (\ref{Expansion_5}), we obtain
\begin{eqnarray}
F \left(z, r, r^{\prime}\right) &=& \frac{4\,\pi}{2}\sum\limits_{l=0}^{\infty}
\sum\limits_{m=-l}^{l}\,Y_{l m} \left(\theta,\phi\right)
\; Y_{l m}^{\ast} \left(\theta^{\prime}, \phi^{\prime}\right)\;
\int\limits_{-1}^{+1} d x\;P_l \left(x\right) \,
F \left(x, r, r^{\prime}\right)\,.
\label{Expansion_20}
\end{eqnarray}

\noindent
Now we use a relation between spherical harmonicals and STF-tensors (Eq.~(2.11) in \cite{Thorne}, or Eq.~(2.19)
in \cite{Hartmann_Soffel_Kioustelidis})
\begin{eqnarray}
Y_{l m} \left(\theta,\phi\right) &=& \hat{Y}^{l m}_{L}\,\hat{n}_{L} \left(\theta,\phi\right)\;,
\label{Expansion_25}
\end{eqnarray}

\noindent
where
\begin{eqnarray}
\hat{n}_{L} \left(\theta,\phi\right) &=& 
\underset{i_1,i_2,\,...\,i_l}{\rm STF}\frac{x_{i_1}}{r}\,\frac{x_{i_2}}{r}\,.\,.\,.\,\frac{x_{i_l}}{r}\,,
\label{Expansion_26}
\end{eqnarray}

\noindent
and $\displaystyle n_x + i\,n_y = e^{i\,\phi}\,\sin \theta\;,n_z = \cos \theta$
(Eq.~(2.10) in \cite{Thorne}). The coefficients $\hat{Y}^{l m}_L$ (given by Eqs.~(A6a) - (A6c)
in \cite{Blanchet_Damour1}, or by Eq.~(2.21) in \cite{Hartmann_Soffel_Kioustelidis})
depend on $l,m$ and on $L$, but they are independent of $\left(\theta,\phi\right)$. Using (\ref{Expansion_25})
we verify 
\begin{eqnarray}
F \left(z, r, r^{\prime}\right) &=& \frac{4\,\pi}{2}\sum\limits_{l=0}^{\infty}
\sum\limits_{m=-l}^{l}\,\hat{Y}^{l m}_{L}\,\hat{n}_{L} \left(\theta,\phi\right)\,
Y_{l m}^{\ast} \left(\theta^{\prime}, \phi^{\prime}\right)\;
\int\limits_{-1}^{+1} d x\,P_l \left(x\right)\, F\left(x, r, r^{\prime}\right).
\label{Expansion_30}
\end{eqnarray}

\noindent
By implementing the inversion of Eq.~(\ref{Expansion_25}) (see Eq.~(2.23) in \cite{Hartmann_Soffel_Kioustelidis})
\begin{eqnarray}
\sum\limits_{m=-l}^{l}\hat{Y}^{l\,m}_L\;Y_{l\,m}^{\ast} \left(\theta^{\prime}, \phi^{\prime}\right) &=&
\frac{\left(2\,l+1\right)!!}{4\,\pi\;l!}\hat{n}_L \left(\theta^{\prime}, \phi^{\prime}\right),
\label{Expansion_35}
\end{eqnarray}

\noindent
we obtain
\begin{eqnarray}
F \left(z, r, r^{\prime}\right) &=& \frac{1}{2}\sum\limits_{l=0}^{\infty} \frac{\left(2 l +1\right)!!}{l!}\;
\hat{n}_L \left(\theta,\phi\right)\,\hat{n}_L \left(\theta^{\prime},\phi^{\prime}\right)
\int\limits_{-1}^{+1} d x\;F\left(x, r, r^{\prime}\right)\;P_l \left(x\right)\,.
\label{Expansion_40}
\end{eqnarray}

This expansion of a function of Hilbert space $V=L^2$ into a series of Legendre polynomials has been given by Eq.~(A.26)
in \cite{Blanchet_Damour1}. According to Eq.~(\ref{Green_Function}), the function $F$ as part of the integrand in 
Eq.~(\ref{Expansion_40}) is given by:
\begin{eqnarray}
F\left(x, r, r^{\prime}\right) &=&
\frac{\delta \left(t^{\prime}-t+\frac{\displaystyle\sqrt{r^2+{r^{\prime}}^2-2\,r\,r^{\prime}\,x}}{\displaystyle c}\right)}
{\sqrt{r^2+{r^{\prime}}^2-2\,r\,r^{\prime}\,x}}\;.
\label{Proof2_10}
\end{eqnarray}

\noindent
Note, that the Green function (\ref{Proof2_10}) is automatically retarded since $t \ge t^{\prime}$. 
Using the formula
\begin{eqnarray}
\delta \left( f \left(x\right) \right) &=& \sum\limits_{i=1}^n 
\frac{\delta \left(x - \nu_i\right)}{\left|f^{\prime} \left(\nu_i\right)\right|}\,,
\label{Proof2_15}
\end{eqnarray}

\noindent
where $\nu_i$ are the roots of $f$, i.e. $f \left(\nu_i\right)=0$, 
and $f^{\prime} \left(\nu_i\right) = \frac{\partial f\left(x\right)}{\partial x}|_{x=\nu_i}$, and taking into account that 
the only root of 
$f\left(x\right) = t^{\prime} - t + \frac{\displaystyle\sqrt{r^2+{r^{\prime}}^2-2\,r\,r^{\prime}\,x}}{\displaystyle c}$
is given by (cf. Eq.~(\ref{Proof_Theorem_20})):
\begin{eqnarray}
\nu &=& \frac{r^2 + {r^{\prime}}^2 - c^2 \left(t - t^{\prime}\right)^2}{2\,r\,r^{\prime}}\;,
\label{Proof2_25}
\end{eqnarray}

\noindent
we can rewrite the function (\ref{Proof2_10}) as follows:
\begin{eqnarray}
F\left(x, r, r^{\prime}\right) &=& \Theta \left(t - t^{\prime}\right)\;\Theta \left(1 - \left|\nu\right|\right)\;
\frac{c\,\delta \left(x - \nu\right)}{r\,r^{\prime}}\;.
\label{Proof2_20}
\end{eqnarray}

\noindent
Here, by the Heaviside function $\Theta \left(t - t^{\prime}\right)$ we have taken into account the fact that the
Green function (\ref{Proof2_10}) is retarded, i.e. $t > t^{\prime}$.
Moreover, since the root $\nu$ in (\ref{Proof2_25}) can take arbitrarily large numerical values, we have to
consider the fact that $-1 \le x \le + 1$, which is taken into account by the Heaviside function
$\Theta \left(1 - \left|\nu\right|\right)$ in (\ref{Proof2_20}).
By inserting relation (\ref{Proof2_20}) into Eq.~(\ref{Expansion_40}) we can calculate the integral and get 
\begin{eqnarray}
F \left(\ve{n} \cdot \ve{n}^{\prime}, r, r^{\prime}\right) = \frac{c}{2\;r\,r^{\prime}}\,\Theta \left(t - t^{\prime}\right)
\Theta \left(1 - \left|\nu\right|\right)\,\sum\limits_{l=0}^{\infty}
\frac{\left(2 l +1\right)!!}{l!}\;\hat{n}_L \left(\theta,\phi\right)\,\hat{n}_L \left(\theta^{\prime},\phi^{\prime}\right)
P_l \left(\nu\right)\,.
\label{Proof2_30}
\end{eqnarray}

\noindent
This result is in agreement with Eq.~(\ref{Proof_Theorem_15}). {\bf q.e.d.}
}

\vspace{1.0cm}
Furthermore, the source $T \left(t^{\prime},\ve{x}^{\prime}\right)$ in (\ref{Proof_Theorem_10}) is expanded in spherical 
harmonics (cf. Eq.~(B.4) in \cite{Blanchet_Damour2}), which means in STF notation:
\begin{eqnarray}
T \left(t^{\prime},\ve{x}^{\prime}\right) &=& \sum\limits_{l=0}^{\infty} 
\hat{n}_L \left(\phi^{\prime}\,,\,\theta^{\prime}\right)\;\hat{T}_L \left(t^{\prime}, r^{\prime}\right),
\label{Proof_Theorem_25}
\end{eqnarray}

\noindent
where $\hat{T}_L$ are some STF tensorial functions, but their explicit structure is not relevant here because 
later the inversion of (\ref{Proof_Theorem_25}) will be used, see Eq.~(\ref{Proof_Theorem_125}). 
Inserting the expansions (\ref{Proof_Theorem_15}) and (\ref{Proof_Theorem_25}) into (\ref{Proof_Theorem_10}) yields 
(in spherical coordinates we have 
$d^3 x^{\prime} = dr^{\prime} {r^{\prime}}^2\,\sin \theta^{\prime} d \theta^{\prime}\,d \phi^{\prime}$ and 
$r = \left|\ve{x}\right|$, $r^{\prime} = \left|\ve{x}^{\prime}\right|$):
\begin{eqnarray}
\overline{h} \left(t,\ve{x}\right) &=& \frac{c}{2\,r}\,\int\limits_{- \infty}^{t} dt^{\prime}\int_V d r^{\prime}\;r^{\prime} 
\sum\limits_{l=0}^{\infty}\;\hat{T}_L \left(t^{\prime},r^{\prime}\right)\;
\Theta \left(t - t^{\prime}\right) \;\Theta \left( 1 - \left| \nu \right|\right)
\nonumber\\
\nonumber\\
&& \hspace{-1.5cm} \times \sum\limits_{q=0}^{\infty} \frac{ \left(2\,q + 1\right)!!}{q!}\;P_q \left(\nu\right)
\hat{n}_Q\left(\phi\,,\,\theta\right)\;
\int\limits_{0}^{2 \pi} d \theta^{\prime}\,\sin \theta^{\prime} \int \limits_0^{\pi} d \phi^{\prime}\; 
\hat{n}_L \left(\phi^{\prime}\,,\,\theta^{\prime}\right) \;
\hat{n}_Q \left(\phi^{\prime}\,,\,\theta^{\prime}\right)\,.
\label{Proof_Theorem_30}
\end{eqnarray}

\noindent
For the integration over the angles $\theta^{\prime}$ and $\phi^{\prime}$ we obtain 
(see Eq.~(2.5) in \cite{Thorne}):
\begin{eqnarray}
\int\limits_{0}^{2 \pi} d \theta^{\prime}\,\sin \theta^{\prime} \int \limits_0^{\pi} d \phi^{\prime}\;
\hat{n}_L \left(\phi^{\prime}\,,\,\theta^{\prime}\right) \;
\hat{n}_Q \left(\phi^{\prime}\,,\,\theta^{\prime}\right) &=& \frac{4\,\pi\,l!}{\left(2\,l+1\right)!!}\,\delta_{l\,q}\;.
\label{Proof_Theorem_35}
\end{eqnarray}

\noindent
Thus, we arrive at (cf. Eq.~(D 3) in \cite{Blanchet_Damour1}) 
\begin{eqnarray}
\overline{h} \left(t,\ve{x}\right) &=& \frac{4 \pi\,c}{2 r} \sum\limits_{l=0}^{\infty} \hat{n}_L\left(\phi,\theta\right)
\int\limits_{- \infty}^{t} dt^{\prime} 
\int_V d r^{\prime}\,r^{\prime}\,\hat{T}_L \left(t^{\prime},r^{\prime}\right) P_l \left(\nu\right)
\Theta \left(t - t^{\prime} \right) \Theta \left(1 - \left| \nu \right|\right).
\label{Proof_Theorem_40}
\end{eqnarray}

\noindent
Now we introduce the following four variables (cf. Eqs.~(D4 a) and (D4 b) in \cite{Blanchet_Damour1}) 
which are independent of each other:
\begin{eqnarray}
u &=& c\,t - r \;,\quad u^{\prime} = c t^{\prime} - r^{\prime}\;,
\label{Proof_Theorem_45}
\\
v &=& c\,t + r \;,\quad v^{\prime} = c t^{\prime} + r^{\prime}\;.
\label{Proof_Theorem_50}
\end{eqnarray}

\noindent
After coordinate transformation (\ref{Proof_Theorem_45}) - (\ref{Proof_Theorem_50}), the previous integration domain 
of (\ref{Proof_Theorem_40}), 
${\cal D} = \{ \left( t^{\prime},r^{\prime} \right): - \infty \le t^{\prime} \le t\;{\rm and}\;0\le r^{\prime} < r_0 \}$,
is given by (cf. comments above Eq.~(D5) in \cite{Blanchet_Damour1})
\begin{eqnarray}
{\cal D} = \{ \left( u^{\prime},v^{\prime} \right): u \le v^{\prime} \le v \; {\rm and} \; u^{\prime} \le u \}\;.
\label{Proof_Theorem_55}
\end{eqnarray}

\vspace{1.0cm}

{\footnotesize
{\bf Proof 2:} We will show that the integration domain ${\cal D}$ is given by (\ref{Proof_Theorem_55}).
From the definition of the new variables (\ref{Proof_Theorem_45}) and (\ref{Proof_Theorem_50}) follows
\begin{eqnarray}
c t &=& \frac{u + v}{2} \;,\quad c t^{\prime} = \frac{u^{\prime} + v^{\prime}}{2}\;,
\label{Proof3_5}
\\
r &=& \frac{v - u}{2} \;,\quad r^{\prime} = \frac{v^{\prime} - u^{\prime}}{2}\;.
\label{Proof3_10}
\end{eqnarray}

\noindent
Let us consider the Heaviside function $\Theta \left(1 - \left|\nu\right|\right)$, i.e. the relation:
\begin{eqnarray}
\left| \frac{r^2 + {r^{\prime}}^2 - c^2 \left(t - t^{\prime}\right)^2}{2\,r\,r^{\prime}}\right| &\le& 1 \;.
\label{Proof3_15}
\end{eqnarray}

\noindent
This relation can also be written as  follows:
\begin{eqnarray}
\left(r + r^{\prime}\right)^2 &\ge& c^2 \left(t - t^{\prime}\right)^2\;,
\label{Proof3_20}
\\
\nonumber\\
\left(r - r^{\prime}\right)^2 &\le& c^2 \left(t - t^{\prime}\right)^2\;.
\label{Proof3_25}
\end{eqnarray}

\noindent
First we consider condition (\ref{Proof3_20}). Due to Heaviside function $\Theta \left(t - t^{\prime}\right)$,
i.e. $t \ge t^{\prime}$, we can rewrite (\ref{Proof3_20}) as follows:
\begin{eqnarray}
r + r^{\prime} &\ge& c \left(t - t^{\prime}\right)\;.
\label{Proof3_30}
\end{eqnarray}

\noindent
By inserting (\ref{Proof3_5}) and (\ref{Proof3_10}) into (\ref{Proof3_30}) we find
\begin{eqnarray}
v^{\prime} &\ge& u\;.
\label{Proof3_35}
\end{eqnarray}

\noindent
Now let us consider condition (\ref{Proof3_25}), which can also be written as
\begin{eqnarray}
c \left(t^{\prime} - t \right) &\le& r - r^{\prime} \le c \left(t - t^{\prime}\right)\,.
\label{Proof3_40}
\end{eqnarray}

\noindent
By inserting (\ref{Proof3_5}) and (\ref{Proof3_10}) into (\ref{Proof3_30}) we obtain from both
conditions in (\ref{Proof3_40}):
\begin{eqnarray}
v^{\prime} &\le& v \quad {\rm and} \quad u^{\prime} \le u\;.
\label{Proof3_45}
\end{eqnarray}

\noindent
Finally, the conditions (\ref{Proof3_35}) and (\ref{Proof3_45}) can be summarized as follows:
\begin{eqnarray}
{\cal D} = \{ \left( u^{\prime},v^{\prime} \right): u \le v^{\prime} \le v \; {\rm and} \; u^{\prime} \le u \}\;,
\label{Proof3_50}
\end{eqnarray}

\noindent
which is just the integration domain (\ref{Proof_Theorem_55}). {\bf q.e.d.} 
}

\vspace{1.0cm}

Then, the integral (\ref{Proof_Theorem_40}) in these new variables (\ref{Proof_Theorem_45}) - (\ref{Proof_Theorem_50})
is given by (cf. Eq.~(D5) in \cite{Blanchet_Damour1})
\begin{eqnarray}
\overline{h} \left(t,\ve{x}\right) &=& \frac{4\pi}{4 \left(v-u\right)} 
\sum\limits_{l=0}^{\infty}\,\hat{n}_L\left(\phi,\theta\right)
\int\!\!\!\!\int_{\cal D} du^{\prime}\,dv^{\prime} \left(v^{\prime} - u^{\prime}\right)\,
\hat{T}_L \left(\frac{u^{\prime} + v^{\prime}}{2\,c},\frac{v^{\prime}-u^{\prime}}{2}\right)
\nonumber\\
\nonumber\\
&& \times P_l \left(1 - 2 \frac{\left(u - u^{\prime}\right) \left(v - v^{\prime}\right)}
{\left(v - u\right) \left(v^{\prime} - u^{\prime}\right)}\right).
\label{Proof_Theorem_60}
\end{eqnarray}

\vspace{1.0cm}

{\footnotesize
{\bf Proof 3:} We will show how to arrive at (\ref{Proof_Theorem_60}).
First we note, by means of relations (\ref{Proof3_5}) and (\ref{Proof3_10}), that
\begin{equation}
d t^{\prime}\, d x^{\prime} = \left| \begin{array}[c]{c}
\displaystyle
\frac{\partial t^{\prime}}{\partial u^{\prime}} \quad \frac{\partial t^{\prime}}{\partial v^{\prime}} \\
\nonumber\\
\displaystyle
\frac{\partial x^{\prime}}{\partial u^{\prime}} \quad \frac{\partial x^{\prime}}{\partial v^{\prime}}  \\
\end{array}\right| d u^{\prime}\,d v^{\prime} = \frac{1}{2\,c}\; d u^{\prime}\,d v^{\prime} \,.
\label{Proof4_5}
\end{equation}

\noindent
Then, using $\displaystyle \frac{1}{2\,r} = \frac{1}{v - u}$ and
$\displaystyle r^{\prime} = \frac{v^{\prime} - u^{\prime}}{2}$ from (\ref{Proof3_10}), we obtain from
(\ref{Proof_Theorem_40}) as intermediate step:
\begin{eqnarray}
\overline{h} \left(t,\ve{x}\right) &=& \frac{4\pi}{4\left(v-u\right)} 
\sum\limits_{l=0}^{\infty}\,\hat{n}_L\left(\phi,\theta\right)
\int\!\!\!\!\int_{\cal D} du^{\prime}\,dv^{\prime} \left(v^{\prime} - u^{\prime}\right)\,
\hat{T}_L \left(t^{\prime}, r^{\prime}\right)\,P_l \left(\nu\right)\,,
\label{Proof4_10}
\end{eqnarray}

\noindent
where we also have implemented the integration domain ${\cal D}$ in virtue of (\ref{Proof_Theorem_55}). 
Now, for the both arguments of the function $\hat{T}_L$ we use 
$\displaystyle t^{\prime}= \frac{v^{\prime} + u^{\prime}}{2\,c}$ according to
(\ref{Proof3_5}) and $\displaystyle r^{\prime} = \frac{v^{\prime} - u^{\prime}}{2}$ according to (\ref{Proof3_10})
and obtain a further intermediate step:
\begin{eqnarray}
\overline{h} \left(t,\ve{x}\right) &=& 
\frac{4\pi}{4 \left(v-u\right)} \sum\limits_{l=0}^{\infty}\,\hat{n}_L\left(\phi,\theta\right)
\int\!\!\!\!\int_{\cal D} du^{\prime}\,dv^{\prime} \left(v^{\prime} - u^{\prime}\right)\,
\hat{T}_L \left(\frac{u^{\prime} + v^{\prime}}{2\,c}, \frac{v^{\prime} - u^{\prime}}{2}\right)\,P_l \left(\nu\right)\,.
\label{Proof4_15}
\end{eqnarray}

\noindent
Finally, we have to reexpress the argument $\nu$ of Legendre polynomial $P_l$ in terms of the new variables
$u, v, u^{\prime}, v^{\prime}$. First, from the definition of $\nu$ given by Eq.~(\ref{Proof_Theorem_20}) and
the new variables given by Eqs.~(\ref{Proof3_5}) - (\ref{Proof3_10}) we get:
\begin{eqnarray}
\nu &=&
\frac{\displaystyle \left(\frac{v - u}{2}\right)^2 + \left(\frac{v^{\prime} - u^{\prime}}{2}\right)^2 -
\left(\frac{u + v}{2} - \frac{u^{\prime} + v^{\prime}}{2} \right)^2}
{\displaystyle 2 \left(\frac{v - u}{2}\right)\,\left(\frac{v^{\prime} - u ^{\prime}}{2}\right)}
= 1 - 2 \frac{\left(u - u^{\prime}\right) \left(v - v^{\prime}\right)}
{\left(v - u\right) \left(v^{\prime} - u^{\prime}\right)}\;,
\label{Proof4_20}
\end{eqnarray}

\noindent
which can easily be checked. Thus, inserting (\ref{Proof4_20}) into (\ref{Proof4_15}) yields
\begin{eqnarray}
\overline{h} \left(t,\ve{x}\right) &=& \frac{4\pi}{4\left(v-u\right)} \sum\limits_{l=0}^{\infty}\,
\hat{n}_L\left(\phi,\theta\right)\int\!\!\!\!\int_{\cal D} du^{\prime}\,dv^{\prime} \left(v^{\prime} - u^{\prime}\right)\,
\hat{T}_L \left(\frac{u^{\prime} + v^{\prime}}{2\,c},\frac{v^{\prime}-u^{\prime}}{2}\right)
\nonumber\\
\nonumber\\
&& \times P_l \left(1 - 2 \frac{\left(u - u^{\prime}\right) \left(v - v^{\prime}\right)}
{\left(v - u\right) \left(v^{\prime} - u^{\prime}\right)}\right)\,,
\label{Proof4_25}
\end{eqnarray}

\noindent
which is just in coincidence with expression (\ref{Proof_Theorem_60}). {\bf q.e.d.} 
}

\vspace{1.0cm}

Then we use the following relation for Legendre polynomial (cf. Eq.~(D6) in \cite{Blanchet_Damour1}):
\begin{eqnarray}
P_l \left(1 - 2 \frac{\left(u - u^{\prime}\right) \left(v - v^{\prime}\right)}
{\left(v - u\right) \left(v^{\prime} - u^{\prime}\right)}\right) &=&
\frac{\left(-1\right)^l}{l!}\,\frac{\left(v - u\right)^{l+1}}{\left(v^{\prime}-u^{\prime}\right)^l}\;
\frac{\partial^l}{\partial u^l}
\left[\frac{\left(u - u^{\prime}\right)^l\,\left(u-v^{\prime}\right)^l}{\left(v-u\right)^{l+1}}\right]\;.
\label{Proof_Theorem_65}
\end{eqnarray}

\vspace{1.0cm}

{\footnotesize
{\bf Proof 4:}
{\it Blanchet} \& {\it Damour} (1986) \cite{Blanchet_Damour1} have found an elegant way to show the validity of 
(\ref{Proof_Theorem_65}) via Euler-Poisson-Darboux differential equation, see text below Eq.~(D 6) 
in \cite{Blanchet_Damour1}. Here, we will demonstrate (\ref{Proof_Theorem_65}) straightaway. 
According to Eq.~(\ref{Proof_Theorem_65}), the Legendre polynomial under consideration is given by
\begin{eqnarray}
P_l \left(\nu\right) &=& P_l \left(1 - 2 \frac{\left(u - u^{\prime}\right) \left(v - v^{\prime}\right)}
{\left(v - u\right) \left(v^{\prime} - u^{\prime}\right)}\right)\,.
\label{Proof11_5}
\end{eqnarray}

\noindent
Using Rodrigues' formula (\ref{Expansion_1}) we verify 
\begin{eqnarray}
P_l \left(\nu\right) = \frac{1}{2^l\,l!}\,\frac{d^l}{d \nu^l} \left(\nu^2 - 1\right)^l
&=& \frac{2^l}{l!}\,\frac{d^l}{d \nu^l}
\left(\frac{\left(u - u^{\prime}\right)^2 \left(v - v^{\prime}\right)^2}{\left(v - u \right)^2
\left(v^{\prime} - u^{\prime}\right)^2} - \frac{\left(u - u^{\prime}\right) \left(v - v^{\prime}\right)}
{\left(v - u\right) \left(v^{\prime} - u^{\prime}\right)}\right)^l
\nonumber\\
\nonumber\\
&=& \frac{2^l}{l!}\,\frac{d^l}{d \nu^l}
\left(\frac{\left(v - v^{\prime}\right)\left(v - u^{\prime}\right)}{\left(v^{\prime} - u^{\prime}\right)^2}
\frac{\left(u - u^{\prime}\right) \left(u - v^{\prime}\right)}{\left(v - u\right)^2}\right)^l \;.
\label{Proof11_10}
\end{eqnarray}

\noindent
Now we use the relation
\begin{eqnarray}
\frac{d}{d \nu} &=& \left(\frac{d u}{d \nu}\right)\;\frac{d}{d u}\,,
\label{Proof11_15}
\end{eqnarray}

\noindent
while for any value of $l$ we obtain
\begin{eqnarray}
\frac{d^l}{d \nu^l} &=&
\underbrace{\left[\left(\frac{d u}{d \nu}\right)\;\frac{\partial}{\partial u}\right]
\times \left[ \left(\frac{d u}{d \nu}\right)\;\frac{\partial}{\partial u}\right]
\times \,.\,.\,. \times \left[\left(\frac{d u}{d \nu}\right)\;\frac{\partial}{\partial u}\right]}_{l}
= \left[\left(\frac{d u}{d \nu}\right)\;\frac{\partial}{\partial u}\right]^l\,.
\label{Proof11_21}
\end{eqnarray}

\noindent
Let us calculate the factor in (\ref{Proof11_15}). For that we have to reconvert 
\begin{eqnarray}
\nu &=& 1 - 2 \frac{\left(u - u^{\prime}\right) \left(v - v^{\prime}\right)}
{\left(v - u\right) \left(v^{\prime} - u^{\prime}\right)}
\label{Proof11_25}
\end{eqnarray}

\noindent
in terms of $u$, and get  
\begin{eqnarray}
u &=& \frac{u^{\prime} + \frac{\displaystyle 1 - \nu}{\displaystyle 2}\,
\frac{\displaystyle v^{\prime}- u^{\prime}}{\displaystyle v - v^{\prime}}\,v}
{1 + \frac{\displaystyle 1 - \nu}{\displaystyle 2}\,
\frac{\displaystyle v^{\prime}- u^{\prime}}{\displaystyle v - v^{\prime}}}\;.
\label{Proof11_30}
\end{eqnarray}

\noindent
With allowance for expression (\ref{Proof11_30}), we have 
\begin{eqnarray}
\left(\frac{d u}{d \nu}\right) &=& \frac{1}{2}\;
\frac{\left(v^{\prime} - u^{\prime}\right) \left(v - u\right)^2}{\left(v^{\prime}-v\right)\left(v - u^{\prime}\right)}\;.
\label{Proof11_35}
\end{eqnarray}

\noindent
In Eq.~(\ref{Proof11_35}), after performing the differentiation, the expression (\ref{Proof11_25}) 
has been reinserted.  
Inserting operator (\ref{Proof11_21}) into (\ref{Proof11_10}), using (\ref{Proof11_35}), yields 
\begin{eqnarray}
P_l \left(\nu\right) = \frac{\left(-1\right)^l}{l!}\,\frac{1}{\left(v^{\prime} - u^{\prime}\right)^l}\;
\left[\left(v-u\right)^2\;\frac{\partial}{\partial u}\right]^l
\left(\frac{\left(u - u^{\prime}\right) \left(u - v^{\prime}\right)}{\left(u - v\right)^2}\right)^l\;.
\label{Proof11_40}
\end{eqnarray}

\noindent
Now we apply the following relation, which is proven in appendix \ref{Relation}:
\begin{eqnarray}
\left[\left(v - u\right)^2 \frac{\partial}{\partial u}\right]^l
\, \frac{\left(u-u^{\prime}\right)^l\left(u-v^{\prime}\right)^l}
{\left(u-v\right)^{2 l}} &=& \left(v-u\right)^{l+1}\,\frac{\partial^l}{\partial u^l}
\frac{\left(u-u^{\prime}\right)^l\left(u-v^{\prime}\right)^l} {\left(v-u\right)^{l+1}}\,.
\label{Proof11_65}
\end{eqnarray}

\noindent
By inserting (\ref{Proof11_65}) into (\ref{Proof11_40}) we obtain
\begin{eqnarray}
P_l \left(\nu\right) = \frac{\left(-1\right)^l}{l!}\,\frac{\left(v-u\right)^{l+1}}{\left(v^{\prime} - u^{\prime}\right)^l}\; \frac{\partial^l}{\partial u^l}
\frac{\left(u-u^{\prime}\right)^l \left(u - v^{\prime}\right)^l}{\left(v-u\right)^{l+1}}\,,
\label{Proof11_70}
\end{eqnarray}

\noindent
which represents the asserted relation (\ref{Proof_Theorem_65}). {\bf q.e.d.}
}

\vspace{1.0cm}

Now, by means of relation (\ref{Proof_Theorem_65}) and with the aid of (cf. Eq.~(A35 b) in \cite{Blanchet_Damour1})
\begin{eqnarray}
\frac{1}{l!}\,\hat{n}_L\left(\theta,\phi\right)\;\left(v - u\right)^l\,\frac{\partial^{2l}}{\partial u^l\,\partial v^l}
\left(\frac{F \left(u\right)}{v - u}\right) &=& \frac{1}{2}\,
\hat{\partial}_L \left(\frac{F \left(c t - r\right)}{r}\right)
\label{Proof_Theorem_70}
\end{eqnarray}

\vspace{1.0cm}

{\footnotesize
{\bf Proof 5:}
We will show the validity of relation (\ref{Proof_Theorem_70}). The function $F$ on the right-hand side in 
Eq.~(\ref{Proof_Theorem_70}) does not depend explicitly on three-vector $\ve{x}$ but only on its absolute value 
$r = \left|\ve{x}\right|$. Therefore, it is meaningful to rewrite the differential operator $\hat{\partial}_L$ in 
a form where the vectorial dependence is projected out of the differential process. This can be achieved with virtue 
of the following relation, see also Eq.~(A.30) in \cite{Blanchet_Damour1}:
\begin{eqnarray}
\hat{\partial}_L \left(\frac{F \left(c t - r\right)}{r}\right) &=& \hat{n}_L\,r^l 
\left(\frac{1}{r}\,\frac{\partial}{\partial r}\right)^l\left(\frac{F \left(c t - r\right)}{r}\right).
\label{Proof10_15}
\end{eqnarray}

\noindent
For proofing (\ref{Proof10_15}) recall
$\frac{\partial f \left(r\right)}{\partial x^{a_1}} = \frac{x^{a_1}}{r}\,\frac{\partial f \left(r\right)}{\partial r}$ and 
one verifies $\partial_L\,f \left(r\right) 
= n_L\,r^l\,\left(\frac{1}{r}\,\frac{\partial}{\partial r}\right)^l\,f\left(r\right)$ plus terms containing at least 
one Kronecker delta which, however, vanish after STF operation; e.g. 
$\underset{ab}{\rm STF}\,\delta_{ab}=0$, $\underset{abc}{\rm STF}\,\delta_{ab}\,x^{c}=0$, etc.

By using the variables $u=ct - r$, $v = ct + r$, see Eqs.~(\ref{Proof_Theorem_45}) and (\ref{Proof_Theorem_50}), we can 
rewrite (\ref{Proof10_15}) as follows:
\begin{eqnarray}
\hat{\partial}_L \left(\frac{F \left(c t - r\right)}{r}\right) &=& 2\,\hat{n}_L\,\left(v-u\right)^l
\left(\frac{1}{v-u}\left(\frac{\partial}{\partial v} - \frac{\partial}{\partial u}\right)\right)^l 
\left(\frac{F \left(u\right)}{v - u}\right), 
\label{Proof10_25}
\end{eqnarray}

\noindent
where we have used the chain rule: 
$\displaystyle \frac{\partial}{\partial r} = \left(\frac{\partial v}{\partial r}\right) 
\frac{\partial}{\partial v} + \left(\frac{\partial u}{\partial r}\right) \frac{\partial}{\partial u} 
= \frac{\partial}{\partial v} - \frac{\partial}{\partial u}$. Now we apply the following identity 
which can easily be proven with the aid of mathematical induction (i.e. show the validity of (\ref{Proof10_30}) 
for $l=1$ and then prove that the validity of (\ref{Proof10_30}) for any one natural number $l$ implies the validity 
of (\ref{Proof10_30}) for the next natural number $l+1$):
\begin{eqnarray}
\left(\frac{1}{v-u}\left(\frac{\partial}{\partial v} - \frac{\partial}{\partial u}\right)\right)^l 
\frac{F \left(u\right)}{v - u} 
&=& \frac{1}{l!}\,\frac{\partial^{2 l}}{\partial u^l\,\partial v^l}\,\frac{F \left(u\right)}{v - u}\,.
\label{Proof10_30}
\end{eqnarray}

\noindent
By inserting (\ref{Proof10_30}) into (\ref{Proof10_25}) we obtain 
\begin{eqnarray}
\frac{1}{2}\, \hat{\partial}_L \left(\frac{F \left(c t - r\right)}{r}\right) &=& \hat{n}_L\;
\frac{\left(v-u\right)^l}{l!}\,\frac{\partial^{2 l}}{\partial v^l \partial u^l}
\left(\frac{F \left(u\right)}{v - u}\right)\,,
\label{Proof10_45}
\end{eqnarray}

\noindent
which is just relation (\ref{Proof_Theorem_70}). {\bf q.e.d.} 
}

\vspace{1.0cm}

we can rewrite Eq.~(\ref{Proof_Theorem_60}) using $\hat{\partial}_L$ as follows (cf. Eq.~(D7) in \cite{Blanchet_Damour1}):
\begin{eqnarray}
\overline{h} \left(t,\ve{x}\right) &=& \sum\limits_{l=0}^{\infty}\frac{\pi}{2}\,\frac{1}{l!}
\int\!\!\!\!\int_{\cal D}\,\frac{du^{\prime}\,dv^{\prime}}{\left(v^{\prime} - u^{\prime}\right)^{l-1}}
\;\hat{T}_L \left(\frac{u^{\prime} + v^{\prime}}{2\,c}, \frac{v^{\prime}-u^{\prime}}{2}\right)
\hat{\partial}_L
\left[\frac{\left(c t - r - u^{\prime}\right)^l\,\left(c t - r - v^{\prime}\right)^l}{r}\right]\;.
\nonumber\\
\label{Proof_Theorem_75}
\end{eqnarray}

\vspace{1.0cm}

{\footnotesize
{\bf Proof 6:}
In order to obtain from Eq.~(\ref{Proof_Theorem_60}) the expression in Eq.~(\ref{Proof_Theorem_75}), the relation
(\ref{Proof_Theorem_65}) is used, which yields
\begin{eqnarray}
\overline{h} \left(t,\ve{x}\right) &=& \frac{4\pi}{4 \left(v-u\right)}\,
\sum\limits_{l=0}^{\infty}\,\hat{n}_L\left(\phi,\theta\right)
\int\!\!\!\!\int_{\cal D} du^{\prime}\,dv^{\prime} \left(v^{\prime} - u^{\prime}\right)\,
\hat{T}_L \left(\frac{u^{\prime} + v^{\prime}}{2\,c},\frac{v^{\prime}-u^{\prime}}{2}\right)
\nonumber\\
\nonumber\\
&& \times \frac{\left(-1\right)^l}{l!}\,\frac{\left(v - u\right)^{l+1}}{\left(v^{\prime}-u^{\prime}\right)^l}\;
\frac{\partial^l}{\partial u^l}
\left[\frac{\left(u - u^{\prime}\right)^l\,\left(u-v^{\prime}\right)^l}{\left(v-u\right)^{l+1}}\right]\;.
\label{Proof6_5}
\end{eqnarray}

\noindent
For being able to apply relation (\ref{Proof_Theorem_70}), we have to rewrite the term
\begin{eqnarray}
\frac{\partial^l}{\partial u^l}
\left[\frac{\left(u - u^{\prime}\right)^l\,\left(u-v^{\prime}\right)^l}{\left(v-u\right)^{l+1}}\right]\,.
\label{Proof6_10}
\end{eqnarray}

\noindent
For doing that, we note the relation
\begin{eqnarray}
\frac{\partial^l}{\partial v^l}
\left[\frac{1}{v - u}\right] &=& \frac{\left(-1\right)^l\,l!}{\left(v-u\right)^{l+1}}\,.
\label{Proof6_15}
\end{eqnarray}

\noindent
By means of this relation we find for the term (\ref{Proof6_10}) the following expression:
\begin{eqnarray}
\frac{\partial^l}{\partial u^l}
\left[\frac{\left(u - u^{\prime}\right)^l\,\left(u-v^{\prime}\right)^l}{\left(v-u\right)^{l+1}}\right]
&=& \frac{\left(-1\right)^l}{l!}\,\frac{\partial^{2l}}{\partial u^l\;\partial v^l} \,
\left[\frac{\left(u - u^{\prime}\right)^l\,\left(u-v^{\prime}\right)^l}{v - u}\right]\,.
\label{Proof6_20}
\end{eqnarray}

\noindent
Inserting relation (\ref{Proof6_20}) into Eq.~(\ref{Proof6_5}) yields 
\begin{eqnarray}
\overline{h} \left(t,\ve{x}\right) &=& \frac{4\pi}{4 \left(v-u\right)}\,
 \sum\limits_{l=0}^{\infty}\,\hat{n}_L\left(\phi,\theta\right)
\int\!\!\!\!\int_{\cal D} du^{\prime}\,dv^{\prime} \left(v^{\prime} - u^{\prime}\right)\,
\hat{T}_L \left(\frac{u^{\prime} + v^{\prime}}{2\,c},\frac{v^{\prime}-u^{\prime}}{2}\right) 
\nonumber\\
\nonumber\\
&& \times
\frac{1}{l!}\,\frac{\left(v - u\right)^{l+1}}{\left(v^{\prime}-u^{\prime}\right)^l}\;
\frac{1}{l!}\,\frac{\partial^{2l}}{\partial u^l\;\partial v^l} \,
\left[\frac{\left(u - u^{\prime}\right)^l\,\left(u-v^{\prime}\right)^l}{v - u}\right]\,,
\label{Proof6_25}
\end{eqnarray}

\noindent
where we have taken into account that $\left(-1\right)^l\,\left(-1\right)^l = 1$.
Now we can apply relation (\ref{Proof_Theorem_70}) and obtain
\begin{eqnarray}
\overline{h} \left(t,\ve{x}\right) &=& \sum\limits_{l=0}^{\infty}\frac{\pi}{2}\,\frac{1}{l!}\,
\int\!\!\!\!\int_{\cal D}\,\frac{du^{\prime}\,dv^{\prime}}{\left(v^{\prime} - u^{\prime}\right)^{l-1}}
\hat{T}_L \left(\frac{u^{\prime} + v^{\prime}}{2\,c}, \frac{v^{\prime}-u^{\prime}}{2}\right)
\hat{\partial}_L
\left[\frac{\left(c t - r - u^{\prime}\right)^l\,\left(c t - r - v^{\prime}\right)^l}{r}\right]\,,
\label{Proof5}
 \end{eqnarray}

\noindent
which is just relation (\ref{Proof_Theorem_75}). {\bf q.e.d.}
}

\vspace{1.0cm}

By means of the transformation 
\begin{eqnarray}
u^{\prime} = s\;,\quad v^{\prime} = s + 2\,y\;,
\label{Coordinate_Transformation_A}
\end{eqnarray}

\noindent
a straightforward calculation shows that (\ref{Proof_Theorem_75}) can be written as follows 
(cf. Eq.~(D9) in \cite{Blanchet_Damour1}):
\begin{eqnarray}
\overline{h} \left(t,\ve{x}\right) &=& \sum\limits_{l=0}^{\infty}\frac{4\,\pi}{2^{l+1} l!} 
\int\limits_{- \infty}^{c t - r} \,d s\,
\int\limits_{\frac{1}{2}\left(c t-r-s\right)}^{\frac{1}{2}\left(ct+r-s\right)} 
\frac{d y}{y^{l-1}}\,\hat{T}_L\left(\frac{s+y}{c}, y\right) 
\hat{\partial}_L
\left[\frac{\left(c t - r - s\right)^l \left(c t - r - s - 2 y\right)^l}
{r}\right] .
\nonumber\\
\label{Proof_Theorem_80}
\end{eqnarray}

\noindent
Furthermore, this expression can be written in the following form (cf. Eq.~(D8) in \cite{Blanchet_Damour1}):
\begin{eqnarray}
\overline{h} \left(t,\ve{x}\right) = - \sum\limits_{l=0}^{\infty}\frac{4\,\pi}{2^{l+1} l!}
\int\limits_{- \infty}^{c t - r} \,d s\,\hat{\partial}_L
&& \Bigg[ \int\limits_{a}^{\frac{1}{2}\left(ct-r-s\right)}
\frac{d y}{y^{l-1}}\,\hat{T}_L \left(\frac{s+y}{c}, y\right) 
\frac{\left(c t - r - s\right)^l \left(c t - r - s - 2 y\right)^l}{r}
\nonumber\\
\nonumber\\
&& \hspace{-3.0cm} - \int\limits_{a}^{\frac{1}{2}\left(ct+r-s\right)}
\frac{d y}{y^{l-1}}\,\hat{T}_L \left(\frac{s+y}{c}, y\right) 
\frac{\left(c t + r - s\right)^l \left(c t + r - s - 2 y\right)^l}{r}\Bigg].
\label{Proof_Theorem_85}
\end{eqnarray}

\noindent
Here, we have commuted the operator $\hat{\partial}_L$ with the integrals, because all differentiations of the
upper limits $\frac{1}{2}\left(ct-\epsilon r-s\right)$ with $\epsilon = \pm 1$ vanish, due to 
the factor $\left(c t - \epsilon\,r - s - 2 y\right)^l$ inside the integrals.

\vspace{1.0cm}

{\footnotesize
{\bf Proof 7:}
We will show how to obtain (\ref{Proof_Theorem_85}) from (\ref{Proof_Theorem_80}). 
First, we separate the second integral in (\ref{Proof_Theorem_80}) into two parts as follows:
\begin{eqnarray}
\overline{h} \left(t,\ve{x}\right) && =  - \sum\limits_{l=0}^{\infty}\frac{4\,\pi}{2^{l+1} l!}\,
\int\limits_{- \infty}^{c t - r} d s \,\int\limits_a^{\frac{1}{2}\left(c t-r-s\right)}
\frac{d y}{y^{l-1}}\;\hat{T}_L \left(\frac{s+y}{c}, y\right)\,\hat{\partial}_L
\left[\frac{\left(c t - r - s\right)^l\,\left(c t - r - s - 2 y\right)^l}{r}\right]
\nonumber\\
\nonumber\\
&& + \sum\limits_{l=0}^{\infty}\frac{4\,\pi}{2^{l+1} l!}\,
\int\limits_{- \infty}^{c t - r} d s \,
\int\limits_{a}^{\frac{1}{2}\left(ct+r-s\right)}
\frac{d y}{y^{l-1}}\;\hat{T}_L \left(\frac{s+y}{c}, y\right)\,
\hat{\partial}_L
\left[\frac{\left(c t - r - s\right)^l\,\left(c t - r - s - 2 y\right)^l}{r}\right]\,,
\label{Proof7_5}
\end{eqnarray}

\noindent
where $a$ is an arbitrarily chosen constant which separates the region of integration variable $y$; in the first line
in (\ref{Proof7_5}) the minus-sign in front of the integral takes into account that we have interchanged the upper and
lower limits of integration. Now let us recall the fundamental theorem of integral calculus:
\begin{eqnarray}
\frac{d}{d x} \int \limits_a^x d y\; f \left(y\right) &=& f \left(y\right)\Bigg|_{y = x} = f \left(x\right)\,.
\label{Proof7_10}
\end{eqnarray}

\noindent
The differential operator $\hat{\partial}_L$ in (\ref{Proof7_5}) contains terms like
$\frac{\partial}{\partial x^k}=\left(\frac{\partial r}{\partial x^k}\right)\frac{\partial}{\partial r}$.
Accordingly, in the first line of (\ref{Proof7_5}) we can take the differential operator $\hat{\partial}_L$
in front of the integral, because the differentiation of the upper limit $\frac{1}{2}\left(c t-r-s\right)$ would yield a
term
\begin{eqnarray}
\left(c t - r - s - 2 y\right)^l\Bigg|_{y = \frac{1}{2}\left(c t-r-s\right)} &=& 0\;,
\label{Proof7_15}
\end{eqnarray}

\noindent
due to the term $\left(c t - r - s - 2 y\right)^l$ in the argument of the integral; the differentiation of the
lower limit $a$ gives zero because $a$ is a constant. Thus, instead of (\ref{Proof7_5}) we can write
\begin{eqnarray}
\overline{h} \left(t,\ve{x}\right) && =  - \sum\limits_{l=0}^{\infty}\frac{4\,\pi}{2^{l+1} l!}\,
\int\limits_{- \infty}^{c t - r} d s\;\hat{\partial}_L \int\limits_a^{\frac{1}{2}\left(c t-r-s\right)}
\frac{d y}{y^{l-1}}\;\hat{T}_L \left(\frac{s+y}{c}, y\right)
\left[\frac{\left(c t - r - s\right)^l\,\left(c t - r - s - 2 y\right)^l}{r}\right]
\nonumber\\
\nonumber\\
&& + \sum\limits_{l=0}^{\infty}\frac{4\,\pi}{2^{l+1} l!} \int\limits_{- \infty}^{c t - r} d s
\int\limits_{a}^{\frac{1}{2}\left(ct+r-s\right)}\frac{d y}{y^{l-1}}\,\hat{T}_L\left(\frac{s+y}{c},y\right)\,\hat{\partial}_L
\left[\frac{\left(c t - r - s\right)^l\,\left(c t - r - s - 2 y\right)^l}{r}\right]\,.
\label{Proof7_20}
\end{eqnarray}

\noindent
Now let us consider the second line in (\ref{Proof7_20}), especially the term:
\begin{eqnarray}
\hat{\partial}_L \left[\frac{\left(c t - r - s\right)^l\,\left(c t - r - s - 2 y\right)^l}{r}\right] = \hat{\partial}_L
\left[\frac{\left[\left(ct - r\right)^2-2\left(s + y\right)\left(ct - r\right) + s\left(s + 2 y\right)\right]^l}{r}\right]\,.
\label{Proof7_25}
\end{eqnarray}

\noindent
We apply the following relation (for a proof of relation (\ref{Proof7_30}) see Appendix \ref{Appendix1},
see also Eq.~(A 36) in \cite{Blanchet_Damour1}):
\begin{eqnarray}
\hat{\partial}_L \left[\frac{\left(c t - r\right)^i}{r}\right] &=&
\hat{\partial}_L \left[\frac{\left(c t + r\right)^i}{r}\right]
\quad {\rm if} \quad i = 0,1,...,2 l\;,
\label{Proof7_30}
\end{eqnarray}

\noindent
and obtain for (\ref{Proof7_25}) the expression
\begin{eqnarray}
\hat{\partial}_L \left[\frac{\left(c t - r - s\right)^l\,\left(c t - r - s - 2 y\right)^l}{r}\right]
&=& \hat{\partial}_L
\left[\frac{\left[\left(ct + r\right)^2-2\left(s + y\right)\left(ct + r\right) + s\left(s + 2 y\right)\right]^l}{r}\right] 
\nonumber\\
\nonumber\\
&=& \hat{\partial}_L \left[\frac{\left(c t + r - s\right)^l\,\left(c t + r - s - 2 y\right)^l}{r}\right]\,.
\label{Proof7_40}
\end{eqnarray}

\noindent
Inserting (\ref{Proof7_40}) into (\ref{Proof7_20}) yields
\begin{eqnarray}
\overline{h} \left(t,\ve{x}\right) && =  - \sum\limits_{l=0}^{\infty}\frac{4\,\pi}{2^{l+1} l!}\,
\int\limits_{- \infty}^{c t - r} d s\;\hat{\partial}_L \int\limits_a^{\frac{1}{2}\left(c t-r-s\right)}
\frac{d y}{y^{l-1}}\;\hat{T}_L \left(\frac{s+y}{c}, y\right)
\left[\frac{\left(c t - r - s\right)^l\,\left(c t - r - s - 2 y\right)^l}{r}\right]
\nonumber\\
\nonumber\\
&& + \sum\limits_{l=0}^{\infty}\frac{4\,\pi}{2^{l+1} l!}\,\int\limits_{- \infty}^{c t - r} d s \,
\int\limits_{a}^{\frac{1}{2}\left(ct+r-s\right)}\frac{d y}{y^{l-1}}\,\hat{T}_L\left(\frac{s+y}{c},y\right)\,\hat{\partial}_L
\left[\frac{\left(c t + r - s\right)^l \left(c t + r - s - 2 y\right)^l}{r}\right]\,.
\label{Proof7_45}
\end{eqnarray}

\noindent
Now, also for the second line of (\ref{Proof7_45}) we can take the differential operator $\hat{\partial}_L$ in front of
the integral, because the differentiation of the upper limit yields terms like
\begin{eqnarray}
\left(c t + r - s - 2 y\right)^l\Bigg|_{y = \frac{1}{2}\left(c t + r - s\right)} &=& 0\;.
\label{Proof7_50}
\end{eqnarray}

\noindent
Accordingly, (\ref{Proof7_45}) can be written as follows:
\begin{eqnarray}
\overline{h} \left(t,\ve{x}\right) && =  - \sum\limits_{l=0}^{\infty}\frac{4\,\pi}{2^{l+1} l!}\,
\int\limits_{- \infty}^{c t - r} d s\;\hat{\partial}_L \int\limits_a^{\frac{1}{2}\left(c t-r-s\right)}
\frac{d y}{y^{l-1}}\;\hat{T}_L \left(\frac{s+y}{c}, y\right)
\left[\frac{\left(c t - r - s\right)^l\,\left(c t - r - s - 2 y\right)^l}{r}\right]
\nonumber\\
\nonumber\\
&& + \sum\limits_{l=0}^{\infty}\frac{4\,\pi}{2^{l+1} l!} \int\limits_{- \infty}^{c t - r} d s \;
\hat{\partial}_L\int\limits_{a}^{\frac{1}{2}\left(ct+r-s\right)}\frac{d y}{y^{l-1}}\;\hat{T}_L \left(\frac{s+y}{c}, y\right)
\left[\frac{\left(c t + r - s\right)^l\,\left(c t + r - s - 2 y\right)^l}{r}\right]\,,
\label{Proof7_55}
\end{eqnarray}

\noindent
which is nothing else but relation (\ref{Proof_Theorem_85}). {\bf q.e.d.}
}

\vspace{1.0cm}

It is readily seen that the expression (\ref{Proof_Theorem_85}) can also be written as 
(cf. Eq.~(6.4) in \cite{Blanchet_Damour1}) 
\begin{eqnarray}
\overline{h} \left(t,\ve{x}\right) &=& - 4\,\pi 
\sum\limits_{l=0}^{\infty} \int\limits_{-\infty}^{c t - r} d s\;\hat{\partial}_L 
\left[\frac{R_a\left(\frac{1}{2}\left(ct-r-s\right),s\right)-R_a\left(\frac{1}{2}\left(ct+r-s\right),s\right) }{r}\right]\,,
\label{Proof_Theorem_95}
\end{eqnarray}

\noindent
with the function (cf. Eq.~(B.6) in \cite{Blanchet_Damour2})
\begin{eqnarray}
R_a \left(r,s\right) &=& r^l \int\limits_a^r d y\, \frac{\left(r - y\right)^l}{l!}\left(\frac{2}{y}\right)^{l-1}\,
\hat{T}_L \left(\frac{s+y}{c},y\right)\,.
\label{Proof_Theorem_100}
\end{eqnarray}

\noindent
Furthermore, by commuting the differential operator $\hat{\partial}_L$ with the integral of the first term in 
(\ref{Proof_Theorem_95}) (for the proof one can use the very same arguments as presented in detail in Proof 7), 
the expression (\ref{Proof_Theorem_95}) can be written as follows 
(see Eq.~(6.8) in \cite{Blanchet_Damour1} or see Eq.~(B.5) in \cite{Blanchet_Damour2}):
\begin{eqnarray}
\overline{h} \left(t,\ve{x}\right) &=& - \sum\limits_{l=0}^{\infty}\hat{\partial}_L 
\Bigg[\frac{4\pi}{r} \int\limits_{-\infty}^{c t - r} \,d s\,R_a\left(\frac{ct-r-s}{2},s\right)\Bigg]
- \sum\limits_{l=0}^{\infty}\int\limits_{-\infty}^{c t - r} \,d s\, \hat{\partial}_L 
\Bigg[\frac{4\pi}{r} R_a\left(\frac{ct+r-s}{2},s\right)\Bigg] .
\nonumber\\
\label{Proof_Theorem_105}
\end{eqnarray}

\noindent
The solution (\ref{Proof_Theorem_105}) is independent of the choice of $a$. In case of a field point outside the 
source $r > r_0$, the first argument of the second term in (\ref{Proof_Theorem_105}) will satisfy 
$\displaystyle \frac{1}{2}\left(ct + r - s\right) > r_0$. 
Hence, if we choose $a = r_0$ it becomes evident from Eq.~(\ref{Proof_Theorem_100}) that the second term in 
(\ref{Proof_Theorem_105}) will vanish when $r > r_0$, because the source is spatially compact (cf. Eq.~(\ref{compact_0}); 
note that the sequence of transformations (\ref{Proof_Theorem_45}), (\ref{Proof_Theorem_50}) and 
(\ref{Coordinate_Transformation_A}) yields $y = \left|\ve{x}^{\prime}\right|$): 
\begin{eqnarray}
\hat{T}_L \left(\frac{s+y}{c},y\right) &=& 0 \quad {\rm for} \; y > r_0\;.
\label{compact_1}
\end{eqnarray}

\noindent
This argumentation immediately yields (Eq.~(B.7) in \cite{Blanchet_Damour2})
\begin{eqnarray}
\overline{h} \left(t,\ve{x}\right) &=& \sum\limits_{l=0}^{\infty} \frac{\left(-1\right)^l}{l!}\;\hat{\partial}_L\,
\left[\frac{\hat{F}_L \left(u\right)}{r}\right]\;,
\label{Proof_Theorem_110}
\end{eqnarray}

\noindent
with (cf. Eq.~(B.8) in \cite{Blanchet_Damour2})
\begin{eqnarray}
\hat{F}_L \left(u\right) &=& 
- 4 \pi \left(-1\right)^l \int\limits_{-\infty}^u d s \left(\frac{u - s}{2}\right)^l 
\int\limits_{r_0}^{\left(u - s\right)/2} dy \left(\frac{u - s}{2} - y\right)^l 
\left(\frac{2}{y}\right)^{l-1}\,\hat{T}_L \left(\frac{s + y}{c}, y\right).
\nonumber\\
\label{Proof_Theorem_115}
\end{eqnarray}

\noindent
By a change of variables $s = u + \left(z - 1\right) y$, the expression (\ref{Proof_Theorem_115}) can be transformed into 
(cf. Eq.~(B.9) in \cite{Blanchet_Damour2}):
\begin{eqnarray}
\hat{F}_L \left(u\right) &=& \frac{4 \pi}{2^{l+1}}\int\limits_{-1}^{+1} d z \left(1 - z^2\right)^l
\int\limits_{0}^{r_0} d y\,y^{l+2}\;\hat{T}_L \left(\frac{u+z\,y}{c},y\right)\;.
\label{Proof_Theorem_120}
\end{eqnarray}

\vspace{1.0cm}

{\footnotesize

{\bf Proof 8:}
We will show the validity of Eq.~(\ref{Proof_Theorem_120}). Consider the expression given by Eq.~(\ref{Proof_Theorem_115}):
\begin{eqnarray}
\hat{F}_L \left(u\right) &=& - 4 \pi 
\left(-1\right)^l \,\int\limits_{-\infty}^u d s \left(\frac{u - s}{2}\right)^l
\int\limits_{r_0}^{\left(u - s\right)/2} dy \left(\frac{u - s}{2} - y\right)^l
\left(\frac{2}{y}\right)^{l-1}\,\hat{T}_L \left(\frac{s + y}{c}, y\right).
\label{Proof9_5}
\end{eqnarray}

\noindent
The transformation reads
\begin{eqnarray}
s &=& u + \left(z - 1\right) y\,,
\label{Proof9_10}
\end{eqnarray}

\noindent
and the differentials 
\begin{equation}
d s\, d y = \left| \begin{array}[c]{c}
\displaystyle
\frac{\partial s}{\partial z} \quad \frac{\partial s}{\partial y} \\
\nonumber\\
\displaystyle
\frac{\partial y}{\partial z} \quad \frac{\partial y}{\partial y}  \\
\end{array}\right|  d z\,d y = y\;d z\,d y \,.
\label{Proof9_15}
\end{equation}

\noindent
Thus, one obtains
\begin{eqnarray}
\hat{F}_L \left(u\right) &=& - \frac{4 \pi}{2^{l+1}}\int\limits_{-\infty}^{u} d z \left(1 - z^2\right)^l
\int\limits_{r_0}^{\left(u-s\right)/2} d y\,y^{l+2}\;\hat{T}_L \left(\frac{u+z\,y}{c}, y\right)\;.
\label{Proof9_25}
\end{eqnarray}

\noindent
Now we have to transform the integration limits. First, we take into account that 
(cf. Eqs.~(\ref{compact_0}) and (\ref{compact_1}))
\begin{eqnarray}
\hat{T}_L \left(\frac{u+z\,y}{c},y\right) &=& 0 \quad {\rm for} \quad y > r_0\;.
\label{Proof9_30}
\end{eqnarray}

\noindent
Consequently, we conclude
\begin{eqnarray}
y_{\rm min} &=& \frac{u-s}{2} \;,\quad y_{\rm max} = r_0\;,
\label{Proof9_35}
\end{eqnarray}

\noindent
and write (\ref{Proof9_25}) as follows:
\begin{eqnarray}
\hat{F}_L \left(u\right) &=& \frac{4 \pi}{2^{l+1}}\,
\int\limits_{z_{\rm min}}^{z_{\rm max}} d z \left(1 - z^2\right)^l
\int\limits^{r_0}_{\left(u-s\right)/2} d y\,y^{l+2}\;\hat{T}_L \left(\frac{u+z\,y}{c}, y\right)\;.
\label{Proof9_45}
\end{eqnarray}

\noindent
From (\ref{Proof9_10}) we conclude  
\begin{eqnarray}
z_{\rm min} &=& \frac{s_{\rm min} - u}{y_{\rm max}} + 1 \;,\quad z_{\rm max} = \frac{s_{\rm max} - u}{y_{\rm min}} + 1 \;.
\label{Proof9_46}
\end{eqnarray}

\noindent
From (\ref{Proof9_35}) and taking into account the upper limit in (\ref{Proof9_25}), i.e. $s \le u$, we immediately get 
\begin{eqnarray}
s_{\rm min} &=& u - 2\,r_0\;,\quad s_{\rm max} = u\;.
\label{Proof9_47}
\end{eqnarray}

\noindent
Then, by inserting (\ref{Proof9_35}) and (\ref{Proof9_47}) into (\ref{Proof9_46}), we obtain the limits:
\begin{eqnarray}
z_{\rm min} &=& -1\;,\quad z_{\rm max} = + 1\;.
\label{Proof9_50}
\end{eqnarray}

\noindent
Accordingly, the integral (\ref{Proof9_45}) reads
\begin{eqnarray}
\hat{F}_L \left(u\right) &=& \frac{4\,\pi}{2^{l+1}} \int\limits_{-1}^{+1} d z \left(1 - z^2\right)^l 
\int\limits_{0}^{r_0} d y\,y^{l+2}\;\hat{T}_L \left(\frac{u+z\,y}{c}, y\right)\,,
\label{Proof9_70}
\end{eqnarray}

\noindent
which is just in coincidence with Eq.~(\ref{Proof_Theorem_120}). {\bf q.e.d.} 
}

\vspace{1.0cm}

Finally, we use the inversion of Eq.~(\ref{Proof_Theorem_25}) (see Eq.~(A9 b) in \cite{Blanchet_Damour1} 
or Eq.~(B.10) in \cite{Blanchet_Damour2})
\begin{eqnarray}
\hat{T}_L \left(t,y\right) &=& \frac{\left(2 l + 1\right)!!}{4\,\pi\,l!} \int\limits_0^{2\,\pi} \sin \theta \, d \theta
\int\limits_0^{\pi} d \phi\; \hat{n}_L \left(\theta,\phi\right) T \left(t,y,\theta,\phi\right).
\label{Proof_Theorem_125}
\end{eqnarray}

\noindent
Inserting (\ref{Proof_Theorem_125}) into (\ref{Proof_Theorem_120}), yields 
for (\ref{Proof_Theorem_110}) the following expression (cf. Eq.~(B.2) in \cite{Blanchet_Damour2})
\begin{eqnarray}
\overline{h} \left(t,\ve{x}\right) &=& \sum \limits_{l=0}^{\infty}  \frac{\left(-1\right)^l}{l!}\,
\partial_L\,\left[\frac{\hat{F}_L \left(u\right)}{r}\right]\;,
\label{Proof_Theorem_130}
\end{eqnarray}

\noindent
where $r = \left|\ve{x}\right|$ is the spatial distance between the origin of coordinate system and the field point. By a 
transformation from spherical coordinates $\ve{y} = \left(y,\theta,\phi\right)$ to Cartesian-like coordinates 
$\ve{x}^{\prime} = \left(x_1^{\prime}, x_2^{\prime}, x_3^{\prime}\right)$, 
the symmetric and tracefree multipole moments of the source are given by  
\begin{eqnarray}
\hat{F}_L \left(u\right) &=& \int_V d^3 x^{\prime} \;\hat{x}_L^{\prime} \int \limits_{-1}^{+1} d z\;\delta_l \left(z\right)\;
T \left(\frac{u + z\,r^{\prime}}{c}, \ve{x}^{\prime}\right)\,,
\label{Proof_Theorem_140}
\end{eqnarray}

\noindent
where the spatial integral runs over the volume $V$ of the source, $r^{\prime} = \left|\ve{x}^{\prime}\right|$ is the 
spatial distance between the origin of coordinate system and a point inside the source with spatial coordinate 
$\ve{x}^{\prime}$, and $u = c t - r$, cf. Eq.~(\ref{Proof_Theorem_45}).
In order to derive the form of Eq.~(\ref{Proof_Theorem_130}), we also have used the relation 
$\hat{\partial}_L \hat{F}_L \left(u\right) = \partial_L \hat{F}_L \left(u\right)$ since $\hat{F}_L$ are STF multipoles, 
that means the trace over any pair of indices in $\hat{F}_L$ vanishes: e.g. for $l=2$ we would have 
$\displaystyle \hat{\partial}_{i_1 i_2}=\frac{\partial^2}{\partial x^{i_1}\,\partial x^{i_2}} 
- \frac{\delta_{i_1 i_2}}{3}\,\frac{\partial^2}{\partial r^2}$,
and due to $\delta_{i_1 i_2}\,\hat{F}_{i_1 i_2}=0$, 
we have $\hat{\partial}_{i_1 i_2}\,\hat{F}_{i_1 i_2} = \partial_{i_1 i_2}\,\hat{F}_{i_1 i_2}$, and so on. 

The functions in (\ref{Proof_Theorem_140}) are given by
\begin{eqnarray}
\delta_l (z) &=& \frac{\left(2\,l + 1\right)!!}{2^{l+1}\,l!}\;\left(1 - z^2\right)^l\;.
\label{Proof_Theorem_145}
\end{eqnarray}

\noindent
In view that $\overline{h}$ in (\ref{Proof_Theorem_130}) stands either for $\overline{h}^{0 0}$, $\overline{h}^{0 i}$, 
or $\overline{h}^{i j}$, while $T$ in (\ref{Proof_Theorem_140}) stands either for 
$\frac{\displaystyle 4\,G}{\displaystyle c^4}\,T^{00}$, $\frac{\displaystyle 4\,G}{\displaystyle c^4}\,T^{0i}$, or 
$\frac{\displaystyle 4\,G}{\displaystyle c^4}\,T^{ij}$, respectively, we can rewrite Eq.~(\ref{Proof_Theorem_130}) and 
(\ref{Proof_Theorem_140}) in terms of their explicit tensorial structure: 
\begin{eqnarray}
\overline{h}^{\alpha \beta} \left(t , \ve{x}\right) &=&
\frac{4\,G}{c^4}\;\sum\limits_{l=0}^{\infty} \frac{\left(-1\right)^l}{l!}\,\partial_L \,
\left[\frac{\hat{F}_L^{\alpha \beta} \left(u\right)}{r}\right]\,,
\label{Proof_Theorem_150}
\end{eqnarray}

\noindent
where the STF multipoles are given by  
\begin{eqnarray}
\hat{F}_L^{\alpha \beta} \left(u\right) &=& \int_V d^3 x^{\prime}\;\hat{x}_L^{\prime} \int\limits_{-1}^{+1} d z\;
\delta_l \left(z\right)\;T^{\alpha \beta} \left(\frac{u + z\,r^{\prime}}{c}, \ve{x}^{\prime}\right)\,.
\label{Proof_Theorem_155}
\end{eqnarray}

\noindent
The equations (\ref{Proof_Theorem_150}) and (\ref{Proof_Theorem_155}) represent the fundamental theorem of STF multipole 
expansion in post-Minkowskian approximation, as previously emphasized by 
Eqs.~(\ref{Introduction_3}) - (\ref{Introduction_4}) in the introductory section. 
In virtue of equation (\ref{compact_0})
(compact support source) it is obvious that the multipole expansion (\ref{Proof_Theorem_150}) and 
(\ref{Proof_Theorem_155})
is valid for regions $r > r_0$, where
$r_0$ is the radius of the smallest possible sphere which encloses completely
the matter source. Finally, it should be noted that a straightforward application of
theorem (\ref{Proof_Theorem_130}) and (\ref{Proof_Theorem_140})
for the case of electrodynamics leads immediately to the STF expansion given by
Eqs.~(4.2) and (4.3) in \cite{Multipole_Damour_2}.

\section{Summary}\label{section_D}

In linearized gravity the Einsteins field equations are given by an inhomogeneous partial differential equation 
(\ref{Introduction_1}) for each of the $10$ components of the metric tensor. In the region exterior to the source 
the retarded solution (\ref{Introduction_2}) can be expanded in terms of $10$ Cartesian STF 
multipoles in post-Minkowskian approximation: Eqs.~(\ref{Introduction_3}) - (\ref{Introduction_4}) 
(= Eqs.~(\ref{Proof_Theorem_150}) and (\ref{Proof_Theorem_155})). These $10$ multipoles in (\ref{Introduction_4}) 
are not independent of each other, because using energy-momentum conservation (four relations) and gauge transformation 
(four relations) they can be reduced to finally $2$ independent STF multipoles: mass multipoles and spin multipoles, 
$\hat{M}_L$ and $\hat{S}_L$, respectively, in post-Newtonian approximation demonstrated 
by {\it Thorne} (1980) \cite{Thorne} and 
{\it Blanchet} $\&$ {\it Damour} (1986,1989) \cite{Blanchet_Damour1,Blanchet_Damour2}, 
while in post-Minkowskian approximation this fact has been 
established by {\it Damour} $\&$ {\it Iyer} (1991) \cite{Multipole_Damour_2}. 

Meanwhile, the STF multipole expansion has become an important tool in linearized gravity and has 
demonstrated its efficiency for a wide spectrum of applications:  in celestial mechanics 
\cite{Hartmann_Soffel_Kioustelidis,DSX1,DSX2}, in the theory of gravitational waves 
\cite{Eubanks,Radiation_Condition,Gravitational_Waves2}, and in high precision astrometry 
where a particularly important aspect thereof is the theory of light propagation in curved
space-time \cite{Lightpropagation1,Lightpropagation2,Lightpropagation3,Lightpropagation4}.

The theorem (\ref{Introduction_3}) - (\ref{Introduction_4}) is the fundamental theorem and the heart part of STF 
multipole expansion; see Eqs.~(B.2) - (B.3) in \cite{Blanchet_Damour2}, Eqs.~(5.3) - (5.4) in \cite{Multipole_Damour_2}, 
Eqs.~(56) - (57) in \cite{Radiation_Condition}, or Eq.~(25) in \cite{Gravitational_Waves2}. 
But despite its formidable importance, an explicit proof of 
Eqs.~(\ref{Introduction_3}) - (\ref{Introduction_4}) has not been presented so far, while some parts of the 
mathematical proof are distributed into several publications \cite{CMM,Thorne,Blanchet_Damour1,Blanchet_Damour2}. In this  
investigation, a detailed proof of the STF multipole decomposition in form of a more didactical 
manuscript has been presented. Only three and rather weak assumptions are required for the validity of the STF multipole expansion: 
\begin{enumerate}
\item[1.] No-incoming radiation condition, Eq.~(\ref{No_Radiation_Condition}). 
\item[2.] The source is spatially compact, Eq.~(\ref{compact_0}).
\item[3.] A spherical expansion of the metric outside the source is possible, Eq.~(\ref{Proof_Theorem_25}).
\end{enumerate}

\noindent
We hope that our investigation elucidates fundamental aspects of the main theorem of STF multipole expansion 
(\ref{Introduction_3}), where the multipoles in post-Minkowskian approximation are defined by (\ref{Introduction_4}).

\section*{Acknowledgment}

The author thanks for encouragement and enlightening discussions with 
Professor Michael H. Soffel, Professor Sergei A. Klioner, and Professor Ralf Sch\"utzhold. 
The work was supported by the Deutsche Forschungsgemeinschaft (DFG). 

\appendix

\section{Proof of Eq.~(\ref{Proof11_65})}\label{Relation}

Relation (\ref{Proof11_65}) contains only derivatives with respect to variable $u$, and since $u$ and $v$ are independent 
variables, here we can treat $v$ as a constant. Accordingly, we introduce a new variable $x=u-v$ with 
$\displaystyle \frac{\partial}{\partial x} = \frac{\partial}{\partial u}$, and rewrite relation (\ref{Proof11_65}) as 
follows: 
\begin{eqnarray}
\left[x^2\,\frac{\partial}{\partial x}\right]^l\,\frac{\left(x+a\right)^l\,\left(x+b\right)^l}{x^{2\,l}} &=& 
x^{l+1}\,\frac{\partial^l}{\partial x^l}\,\frac{\left(x+a\right)^l\,\left(x+b\right)^l}{x^{l+1}}\,,
\label{Appendix_Relation_5}
\end{eqnarray}

\noindent
where $a=v-u^{\prime}$ and $b=v-v^{\prime}$, and the independent variable $u^{\prime}$ and $v^{\prime}$ are 
also considered as constant quantities. In order to show the validity of relation (\ref{Appendix_Relation_5}) 
we apply the binomial theorem: 
\begin{eqnarray}
\left(x+a\right)^l &=&\sum\limits_{p=0}^{l} 
\, \left( \begin{array}[c]{l}
l \\ 
\displaystyle
p \end{array}\right) \,x^{l-p}\,a^p\,, \quad 
\left(x+b\right)^l =\sum\limits_{q=0}^{l} \, \left( \begin{array}[c]{l}
l \\
\displaystyle
q \end{array}\right) \,x^{l-q}\,b^q\,,
\label{Appendix_Relation_10}
\end{eqnarray}

\noindent
where the binomial coefficients are defined by 
\begin{eqnarray}
\left( \begin{array}[c]{l}
l \\
\displaystyle
p \end{array}\right) = \frac{l!}{\left(l-p\right)!\,p!}\,,\quad 
\left( \begin{array}[c]{l}
l \\
\displaystyle
q \end{array}\right) = \frac{l!}{\left(l-q\right)!\,q!}\,.
\label{binomial_coefficient}
\end{eqnarray}

\noindent 
Inserting (\ref{Appendix_Relation_10}) into (\ref{Appendix_Relation_5}) yields  
\begin{eqnarray}
\sum\limits_{p,q=0}^{l}  \left( \begin{array}[c]{l}
l \\
\displaystyle
p \end{array}\right)  \left( \begin{array}[c]{l}
l \\
\displaystyle
q \end{array}\right)\,a^p\,b^q\,
\left[x^2\,\frac{\partial}{\partial x}\right]^l\,\frac{x^{l-p}\, x^{l-q}}{x^{2\,l}} &=& 
\sum\limits_{p,q=0}^{l} \left( \begin{array}[c]{l}
l \\
\displaystyle
p \end{array}\right)  \left( \begin{array}[c]{l}
l \\
\displaystyle
q \end{array}\right)\,a^p\,b^q\,
x^{l+1}\,\frac{\partial^l}{\partial x^l}\,\frac{x^{l-p}\, x^{l-q}}{x^{l+1}}\,.
\nonumber\\
\label{Appendix_Relation_15}
\end{eqnarray}

\noindent
Let us consider each individual term in (\ref{Appendix_Relation_15}). One can easily show the validity of the following 
both relations by means of mathematical induction:
\begin{eqnarray}
\left[x^2\,\frac{\partial}{\partial x}\right]^l\,\frac{x^{l-p}\, x^{l-q}}{x^{2\,l}} &=& 
\left(- 1\right)^l\,x^{-\left(p+q-l\right)} \prod\limits_{k=0}^{l-1} \left(p+q-k\right),
\label{Appendix_Relation_20}
\\
\nonumber\\
x^{l+1}\,\frac{\partial^l}{\partial x^l}\,\frac{x^{l-p}\, x^{l-q}}{x^{l+1}} &=&
\left(- 1\right)^l\,x^{-\left(p+q-l\right)} \prod\limits_{k=0}^{l-1} \left(p+q-k\right).
\label{Appendix_Relation_25}
\end{eqnarray}

\noindent
Accordingly, we can conclude the following identity for each individual term in (\ref{Appendix_Relation_15}): 
\begin{eqnarray}
\left[x^2\,\frac{\partial}{\partial x}\right]^l\,\frac{x^{l-p}\, x^{l-q}}{x^{2\,l}} &=&
x^{l+1}\,\frac{\partial^l}{\partial x^l}\,\frac{x^{l-p}\, x^{l-q}}{x^{l+1}}\,.
\label{Appendix_Relation_30}
\end{eqnarray}

\noindent
That means, each individual term on the left-hand side in (\ref{Appendix_Relation_15}) coincides with the corresponding 
term on the right-hand side in (\ref{Appendix_Relation_15}). Thus, we have shown the validity of relation 
(\ref{Appendix_Relation_15}) and, therefore, the validity of relation (\ref{Appendix_Relation_5}) and (\ref{Proof11_65}), 
respectively.

\section{Proof of Eq.~(\ref{Proof7_30})}\label{Appendix1}

Let us consider both expressions in (\ref{Proof7_30}), which we write as follows 
(for a proof of relation (\ref{Proof8_5}) see Eqs.~(\ref{Proof10_15}) - (\ref{Proof10_45}), while the proof 
of (\ref{Proof8_10}) is very similar, see also relations (A35 b) and (A36 c) in \cite{Blanchet_Damour1}):
\begin{eqnarray}
\hat{\partial}_L \frac{\left( c t - r \right)^n}{r} &=& \frac{2}{l!}\,\hat{n}_L \left(v - u\right)^l\,
\frac{\partial^{2 l}}{\partial u^l\,\partial v^l}\,\frac{u^n}{v - u}\;,
\label{Proof8_5}
\\
\nonumber\\
\hat{\partial}_L \frac{\left( c t + r \right)^n}{r} &=& \frac{2}{l!}\,\hat{n}_L \left(v - u\right)^l\,
\frac{\partial^{2 l}}{\partial u^l\,\partial v^l}\,\frac{v^n}{v - u}\;,
\label{Proof8_10}
\end{eqnarray}

\noindent
where $u = ct - r$ and $v = ct + r$. By subtraction of (\ref{Proof8_5}) from (\ref{Proof8_10}) one obtains 
\begin{eqnarray}
\hat{\partial}_L \frac{\left( c t + r \right)^n}{r} - \hat{\partial}_L \frac{\left( c t - r \right)^n}{r} &=& 
\frac{2}{l!}\,\hat{n}_L \left(v - u\right)^l\,\frac{\partial^{2 l}}{\partial u^l\,\partial v^l}\,\frac{v^n - u^n}{v - u}\;.
\label{Proof8_15}
\end{eqnarray}

\noindent
Now we recall the generalized version of third binomial theorem, 
\begin{eqnarray}
\frac{v^n - u^n}{v - u} &=& \sum \limits_{j=0}^{n-1} v^{n-j-1}\;u^j\;.
\label{Proof8_20}
\end{eqnarray}

\noindent
Due to $2 l \ge n$, the $\left(2 l\right)^{\rm th}$ derivative of the polynomial in (\ref{Proof8_20}) yields zero:
\begin{eqnarray}
\frac{\partial^{2 l}}{\partial u^l\,\partial v^l}\,\frac{v^n - u^n}{v - u} &=& 
\frac{\partial^{2 l}}{\partial u^l\,\partial v^l} \sum \limits_{j=0}^{n-1} v^{n-j-1}\;u^j = 0\;.
\label{Proof8_25}
\end{eqnarray}

\noindent
Thus, inserting (\ref{Proof8_25}) into (\ref{Proof8_15}) yields  
\begin{eqnarray}
\hat{\partial}_L \frac{\left( c t + r \right)^n}{r} - \hat{\partial}_L \frac{\left( c t - r \right)^n}{r} &=& 0\;,
\label{Proof8_30}
\end{eqnarray}

\noindent
which is just relation (\ref{Proof7_30}).

\end{document}